\documentclass[12pt, draftclsnofoot, onecolumn]{IEEEtran}

\usepackage{graphicx}
\graphicspath{ {./fig/} }
\usepackage{latexsym}
\usepackage{amsfonts}
\usepackage{amssymb}
\usepackage{amsbsy}
\usepackage{amsmath}
\usepackage{multicol}
\usepackage{float}
\usepackage{subfigure}
\usepackage[dvips]{epsfig}

\usepackage{comment}
\usepackage{booktabs}
\usepackage{verbatim} 
\usepackage{textcomp}

\usepackage{fancyhdr}
\usepackage{epsfig}
\usepackage{threeparttable}
\usepackage{epsf,epsfig}
\usepackage{amsthm}
\usepackage[noadjust]{cite}
\usepackage{dsfont}
\usepackage[x11names]{xcolor}
\usepackage{enumerate}
\usepackage{setspace}
\usepackage{bbm}
\usepackage{soul}

\usepackage{enumitem}

\usepackage{array}
\makeatletter
\newcommand{\thickhline}{%
  \noalign {\ifnum 0=`}\fi \hrule height 1pt
  \futurelet \reserved@a \@xhline
}
\newcolumntype{"}{@{\hskip\tabcolsep\vrule width 1pt\hskip\tabcolsep}}
\makeatother

\DeclareMathOperator*{\argmax}{arg\,max}  

\usepackage{amsmath}
\usepackage{algorithm}
\usepackage{algpseudocode}
\algrenewcommand{\algorithmiccomment}[1]{\hskip3em$\triangleright$ #1}

\def\({\left(}
\def\){\right)}
\def\[{\left[}
\def\]{\right]}

\def\bu{\boldsymbol{u}}

\def\bd{\boldsymbol{d}}
\def\bw{\boldsymbol{w}}
\def\tv{v}
\def\tu{u}
\def\td{d}
\def\tH{\theta}

\def\bb{\boldsymbol{b}}

\def\SPM{\mathcal{S}}
\def\hSPM{\mathcal{S}}
\def\R{\mathcal{R}}
\def\hR{\mathcal{R}}
\def\vB{\mathcal{B}}
\def\vD{\mathcal{D}}
\def\vU{\mathcal{U}}
\def\vA{\mathcal{A}}

\def\H{\mathcal{H}}
\def\Hn{\mathcal{H}_{net}}

\def\a{\alpha}
\def\b{\beta}

\def\c{\gamma}

\def\l{\ell}
\def\ha{\alpha}
\def\hb{\beta}
\def\hc{\gamma}
\def\hd{\delta}

\def\ind{\mathds{1}}

\def\stwo{SPM }
\def\stwon{SPM}

\begin{document}

\title{Towards Semantic Communication Protocols: A Probabilistic Logic Perspective}



\author{Sejin Seo, $^\dagger$Jihong Park, $^\ddagger$Seung-Woo Ko, $^\dagger$Jinho Choi, \\$^*$Mehdi Bennis, and Seong-Lyun Kim
\thanks{S. Seo and S.-L. Kim are with Yonsei University, Seoul, Korea (e-mail: \{sjseo, slkim\}@ramo.yonsei.ac.kr). 
$^\dagger$J. Park and $^\dagger$J. Choi are with Deakin University, Geelong, VIC 3220, Australia (e-mail: \{jihong.park, jinho.choi\}@deakin.edu.au).
$^\ddagger$S.-W. Ko is with Inha University, Incheon, Korea (e-mail: swko@inha.ac.kr). $^*$M. Bennis is with Oulu University, Finland (e-mail: mehdi.bennis@oulu.fi). This work has been submitted to the IEEE for possible publication. Copyright may be transferred without notice, after which this version may no longer be accessible.}}


\maketitle

\begin{abstract}
Classical medium access control (MAC) protocols are interpretable, yet their task-agnostic control signaling messages (CMs) are ill-suited for emerging mission-critical applications. By contrast, neural network (NN) based protocol models (NPMs) learn to generate task-specific CMs, but their rationale and impact lack interpretability. To fill this void, in this article we propose, for the first time, a semantic protocol model (SPM) constructed by transforming an NPM into an interpretable symbolic graph written in the probabilistic logic programming language (ProbLog). This transformation is viable by extracting and merging common CMs and their connections while treating the NPM as a CM generator. By extensive simulations, we corroborate that the SPM tightly approximates its original NPM while occupying only $0.02$\% memory. By leveraging its interpretability and memory-efficiency, we demonstrate several SPM-enabled applications such as SPM reconfiguration for collision-avoidance, as well as comparing different SPMs via semantic entropy calculation and storing multiple SPMs to cope with non-stationary environments.
\end{abstract}


\section{Introduction}

Traditionally, cellular medium access control (MAC) protocols have been designed primarily for \emph{general purposes}. To this end, classical MAC protocols have been pre-determined by taking into account all possible cases via extensive experiments and standardization activities~\cite{MAC_Handshake}. While handshaking rules and scheduling policies can partly be manipulated (e.g., grant-free access prioritization~\cite{Grantfree_NOMA_IoT_survey}), their \emph{control signaling messages (CMs)} remain unchanged even when tasks and other environmental characteristics vary over time. By its nature, the effectiveness of classical MAC protocols in 6G has recently been questioned~\cite{6G_E2EtHz_commag}. In fact, emerging 6G applications are often mission-critical under non-stationary environments, such as drones and satellites in a non-terrestrial network (NTN)~\cite{6G_NTN}, autonomous cars in a vehicle-to-everything (V2X) network~\cite{6G_V2X}, and visuo-haptic immersive applications in the metaverse~\cite{6G_Metaverse}. These applications cannot be supported without tightly fine-tuning MAC protocols.


Alternatively, \emph{goal-oriented} and task-specific MAC protocols have recently been proposed, wherein signaling messages emerge naturally from a given environment while conducting a downstream task~\cite{Hoydis_2021_arxiv, Hoydis_2021_commag, Han_2020_commag}. For instance, with two user equipments (UEs) and a base station (BS) as visualized in Fig. 1, M. Mota et al. \cite{Hoydis_2021_arxiv} have shown that uplink CMs (UCMs) and downlink CMs (DCMs) in the control plane can emerge for the task of establishing collision-free data communication in the user plane. The key idea comes from interpreting MAC protocol operations through the lens of multi-agent deep reinforcement learning (MADRL) with cheap talk communication among UE agents through the BS~\cite{foerster2016learning}. In this MAC protocol model, hereafter referred to as a \emph{neural protocol model (NPM)}, a single neural network (NN) represents a single cycle of uplink and downlink operations, as illustrated in Fig. 2. 

Precisely, the NN consists of multiple layers, wherein the input and output correspond to the UEs' current states (i.e., buffer states) and actions (i.e., access or silence), respectively. Meanwhile, the activations of two specific layers imply the UCMs and DCMs, respectively. In this architecture, the forward propagation (FP) of the NN describes the sequence of state-UCM-DCM-action operations, and training the NN is tantamount to iteratively giving rewards or Q-values to such sequences under a given environment. Before training, the NN is randomly initialized, and so are the actions, during which these CMs are meaningless activations. However, as the NN training converges, actions become task-optimal, while the UCMs and DCMs gradually emerge as effective CMs in the given task and environment.

\begin{figure*}[t]
  \centering  
  \includegraphics[width=.85\textwidth]{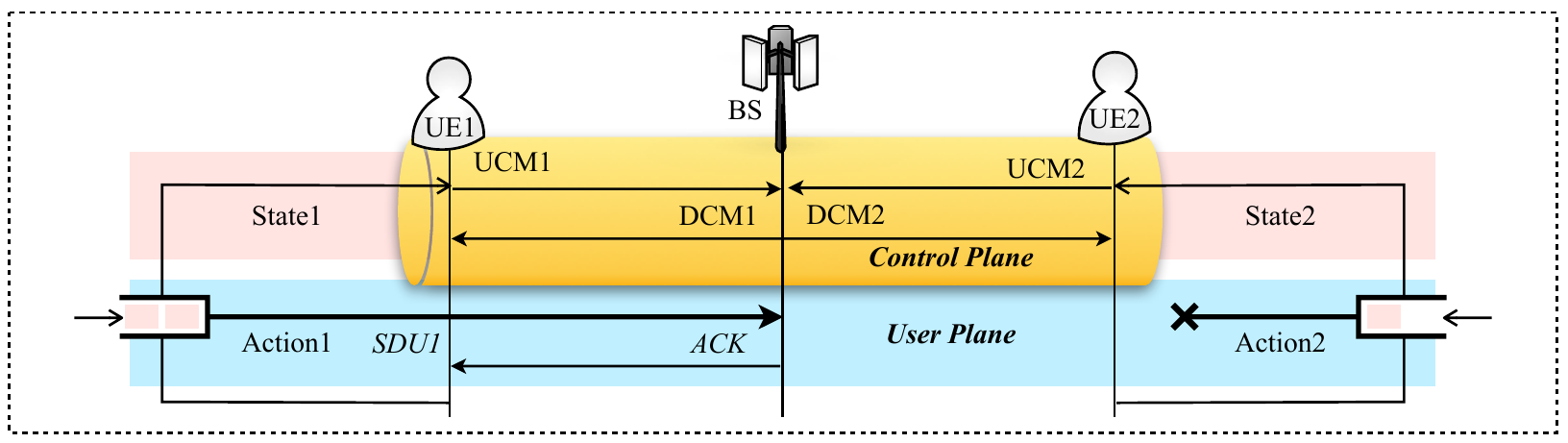}
  \caption{A MAC scenario where two UEs are contending for the same channel by sending UCMs and receiving DCMs. A collision occurs when they send their SDUs simultaneously.}\label{Figure_Cycle}
\end{figure*}

\begin{figure*}[t]
  \centering  
  \includegraphics[width=.9\textwidth]{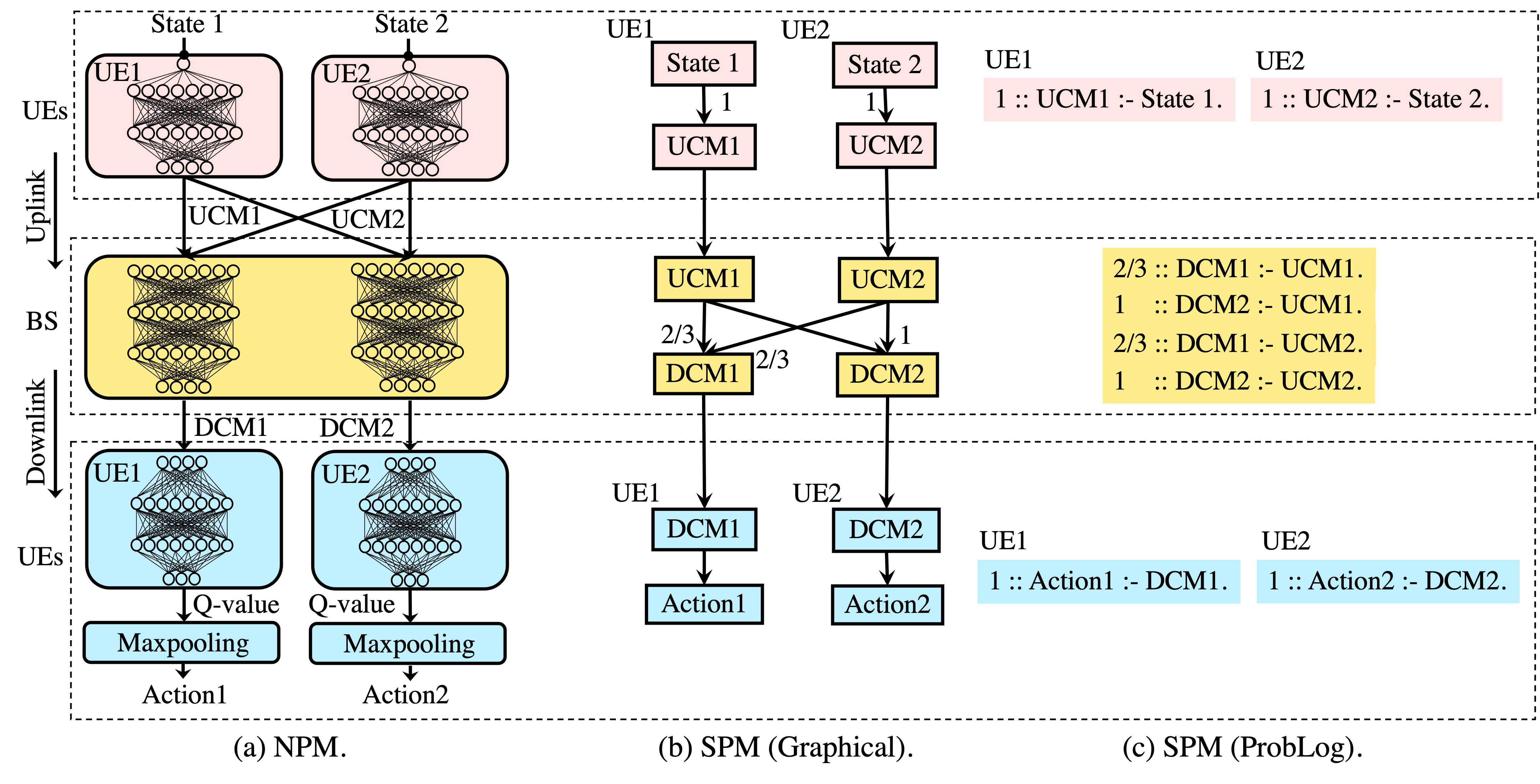}
  \caption{Schematic illustrations and KPI performance of the protocol models: (a) NPM; (b) graphical representation of SPM, and the number on the edges denotes the clause truth probability; and (c) ProbLog representation of SPM.} \label{Fig_Schematics}
\end{figure*}

Despite their effectiveness, NPMs have major disadvantages due to their NN architecture. First, NPMs often overfit to their training samples. By experiment, we observed that even multiple NPMs emerging from the same environmental statistics (e.g., mean traffic arrival rate) share neither common CMs nor their connections (see Sec. \ref{Sec:Disc_Experim}). Second, since the NN architecture is over-parameterized, NPM operations entail large communication payloads, computing time, and memory usage (see Fig. \ref{Fig_Schematics}). Lastly, the NPM is a black-box function where the knowledge regarding the protocol operations is hidden in the NN's model parameters~\cite{achille2019information}, limiting its interpretability or explainability. Besides, we cannot immediately reconfigure the NN's model parameters, because they are updated through iterative gradient descent iterations, degrading its flexibility in manipulating protocol operations.




The overarching goal of this paper is to fill the void between goal-oriented NPMs and general-purpose MAC protocols. As its first step, we revisit the NPM learning scenario considered in~\cite{Hoydis_2021_arxiv}, and propose a novel framework to transform the trained NPM into a \emph{semantic protocol model (SPM)}. This NPM-to-SPM transformation is inspired by human language~\cite{Episodic_Semantic_Memory} and logic programming language for symbolic artificial intelligence (AI)~\cite{Pearl_2019}, as summarized next. At first, treating an NPM (i.e., a trained NN) as the generator of protocol operations, we feed the UE states experienced during training into the NPM, and thereby extract UCMs and DCMs as well as their causal connections. This extraction process is motivated by humans transforming the episodic memory of experiences into the semantic knowledge of general concepts and their relations~\cite{Episodic_Semantic_Memory}. The resultant \emph{NPM extract} shown in Fig. \ref{Fig_ProLog_SPM} is akin to a language for machine agents, and we hereafter refer the UCMs and DCMs as \emph{vocabularies} and their operational connections as \emph{rules}.

\begin{table}[]
\centering
\caption{Comparison of KPIs for NPM and SPM.}
\label{tab:performance}
\resizebox{\textwidth}{!}{%
\begin{tabular}{|c|c|c|c|c|c|c|c|}
\hline
Protocol & \begin{tabular}[c]{@{}c@{}}Goodput\\ (Packets/Cycle)\end{tabular}& \begin{tabular}[c]{@{}c@{}}UCM \\ Vocabularies\end{tabular} & \begin{tabular}[c]{@{}c@{}}DCM \\ Vocabularies\end{tabular} & \begin{tabular}[c]{@{}c@{}}UCM/DCM Length\\ (Bytes)\end{tabular} & \begin{tabular}[c]{@{}c@{}}Model Storage\\ (Bytes)\end{tabular} & \begin{tabular}[c]{@{}c@{}}Inference\\ (FLOPs)\end{tabular} & \begin{tabular}[c]{@{}c@{}}Collision\\ (Collisions/Cycle)\end{tabular}  \\ \hline \hline
NPM & $0.729$ & $10$ & $50$ & $32$ & $4.55$M & $14$K & $0.275$  \\ \hline
SPM & \begin{tabular}[c]{@{}c@{}}$0.729$\\ \textbf{$(100\%)$}\end{tabular} & \begin{tabular}[c]{@{}c@{}}$4$\\ \textbf{$(40\%)$}\end{tabular} & \begin{tabular}[c]{@{}c@{}}$3$\\ \textbf{$(6\%)$}\end{tabular} & \begin{tabular}[c]{@{}c@{}}$0.125$\\ \textbf{$(0.19 \%)$}\end{tabular} & \begin{tabular}[c]{@{}c@{}}$1$K\\\textbf{$(0.02 \%)$}\end{tabular} & \begin{tabular}[c]{@{}c@{}}$8$\\ \textbf{($5.7\times 10^{-4}\%$)}\end{tabular}  & \begin{tabular}[c]{@{}c@{}}$0.0$\\ \textbf{$(0.00\%)$}\end{tabular}  \\ \hline
\end{tabular}%
}
\end{table}



Next, to cope with the limited number of CMs in practice, we reduce the redundant vocabularies and connections in the NPM extract. This is viable by merging vocabularies and connections based on common NN activation patterns and input-output relationships, respectively.
The merging may incur the \emph{problem of polysemy} (a word with multiple meanings), e.g., $A\to B_1\to C_1$ and $A \to B_2 \to C_1$ can be merged into either $A\to B \to C_1$ or $A\to B \to C_2$, incurring the polysemous vocabulary $B$ leading to $C_1$ or $C_2$. Inspired by how humans distinguish the different meanings of a polysemous vocabulary based on communication context~\cite{Polysemy_Jmemlang, Polysemy_ACL}, we make each connection associated with its generation frequency at the NPM as the \textit{context}. 

Finally, we represent each connection of two neighboring vocabularies as a clause written in the \emph{probabilistic logic programming language (ProbLog)}~\cite{problog}, e.g., $A\overset{p}{\to} B$ with probability $p$ is written as $\langle p::A\text{ :- } B \rangle $. Then, the collection of these probabilistic clauses forms an SPM. With this, an SPM is much more than an approximation to its original NPM, providing multi-fold benefits. First, it is symbolic and interpretable, enabling manipulation without re-training, e.g., for immediately controlling collision-avoidance. Second, semantic information stored in each SPM is measurable by evaluating the average entropy of all clauses, enabling the numerical comparison of different SPMs, e.g., for selecting the best SPM for a given environment. Last but not least, as Table \ref{tab:performance} shows, the SPM occupies only 0.02\% of the memory compared to its original NPM. This not only accelerates the protocol operations, but also allows one to store a portfolio of multiple SPMs to cope with a non-stationary environment.

\subsection{Related Works}

The related works aim to design emergent protocols based on NNs, which we call System 1 MAC, however these works overlook the attributes of classical protocols such as reconfigurability and measurability. Also, recent interests in semantic communication and traditional endeavor on symbolic and knowledge-based systems are related to our work, but they are different from SPMs because they either limit their scope to training NN models or focus on point-to-point communication overlooking realistic network-wide communication systems. We acknowledge the works below, which we build upon while articulating our key novelty.

\subsubsection{Classic MAC}
MAC plays a key role in facilitating multiple UEs’ error-free wireless communications via the efficiency-reliability tradeoff~\cite{MAC_Eff_Reliable}. Due to the nature of wireless propagation characteristics, every UE can access the medium whenever they have data to transmit, calling for an efficient utilization of the medium. On the other hand, a UE's excessive access without coordination causes frequent collisions, hampering wireless communication reliability. Based on this tradeoff, various MAC protocols have been proposed for different networks’ specific constraints, e.g., differentiation between primary and secondary users’ priorities for cognitive radios~\cite{Sense-and-Predict}, limited energy and computation capacities of the Internet-of-Things~\cite{IoT_Challenge}, ultra-reliable and ultra-low latency for vehicle-to-everything~\cite{V2X_URLLC} and so on. However, as scenarios and requirements become more complex and diverse, designing a MAC protocol that fulfills those constraints is a daunting task. Moreover, even if this were possible, significant signaling overhead is entailed when implementing the protocol, making it impractical in an actual communication system. 

\subsubsection{Emergent Protocol Design} As explained above, the classical MAC protocol design process is limited because it is not grounded in an actual environment. To cope with such limitation, in system 1 MAC~\cite{Hoydis_2021_arxiv, Hoydis_2021_commag, Han_2020_commag}, the advantage for designing protocols using NN is presented. However, contrary to our endeavor to develop novel semantic communication protocols, most works on emergent communication protocol design are based on learning an NN for a given environment such as wireless MAC using MADRL~\cite{Hoydis_2021_arxiv}, sensor networks, and low-earth orbit satellite networks~\cite{Liu_2006_RLMAC, lee_2021_LEO}. In addition, in~\cite{Pasandi_2020_DeepMAC}, the authors apply NN for optimizing the MAC pipeline by selecting appropriate protocols. However, the performance is bounded by classical protocols that lack grounding in the actual environment. 

\subsubsection{Semantic Communication} Aligned with our interest in learning semantic representations that contain meaning for the environment, recent works on semantic communication aim at achieving service-specific goals via NNs or semantic information theory~\cite{NeSy_Saad_2022, Calvanese_2021_SemCom, Weng_2021_SemCom, Jinho_2022_SemCom}. However, contrary to our interest in developing interpretable and reconfigurable semantic representations, these works focus on using NNs to achieve optimal encoding and decoding parameters~\cite{Weng_2021_SemCom}, or quantify semantic information within the message or the communication agent~\cite{Jihong_2021_SemCom, Jinho_2022_SemCom}. 

\subsubsection{Symbolic Artificial Intelligence} Related to our focus on logical relationships, learning logical relationships to represent expert knowledge has been traditionally embodied by the field of expert systems~\cite{Jackson_1989_Expert}, multi-agent systems~\cite{Calegari_2020}, and knowledge query manipulation language~\cite{Finin_1994}. However, while interesting, these works are not grounded in wireless communication systems. Furthermore, due to the known limitations of subsymbolic (NN-based) AI, the interests in symbolic AI, which attempts to discover the causal structure behind reasoning~\cite{Pearl_2019} and data~\cite{Manhaeve_2020}, is currently surging. However, to date, no tangible approach is available for discovering symbolic knowledge in the context of communication protocols. 

\subsection{Contributions and Organization}

This work is the first of its kind to design a novel MAC protocol based on logic programming and symbolic AI, i.e., SPM. Like an NPM, the SPM is task-specific, and at the same time symbolized and interpretable as in classical MAC protocols, thereby achieving communication and memory efficiencies as well as adaptability to non-stationary environments. Our main contributions are summarized as follows.

\begin{enumerate}
\item \textbf{SPM Construction}: We propose a novel method to construct a ProbLog-based SPM from an NN-based NPM, which occupies only $0.02$\% of the NPM memory usage by extracting and merging semantically common vocabularies.
\item \textbf{SPM Reconfiguration for Collision Avoidance}: Without re-training, we demonstrate that an SPM is reconfigurable for collision avoidance by identifying colliding rules and instantly manipulating their connections written in ProbLog.
\item \textbf{Best SPM Selection via Semantic Entropy}: We empirically show that minimizing the average semantic entropy of an SPM (i.e., mean uncertainty of the SPM operations) achieves the highest goodput or equivalently the highest reward, allowing one to select the best SPM in a stationary environment.
\item \textbf{SPM Portfolio for Non-Stationary Environments}:
By exploiting the memory efficiency of SPMs, we propose an SPM portfolio storing a set of SPMs, each of which is the best SPM for a different environment.
\end{enumerate}

The rest of the paper is organized as follows. In Section II, we explain NPM from the perspective of a MAC problem and summarize its limitations. Section III details SPM construction based on ProbLog and shows its advantages over NPM. Section IV presents the significant attributes of SPM and potential applications, including collision avoidance, SPM selection, and SPM portfolio. In Section V, we numerically demonstrate the effectiveness of SPM by comparing it with several benchmarks. Last, we conclude the work in Section IV.

\section{Revisiting Neural Protocol Learning}

Recognizing the limitations of hand-crafted protocols, we revisit neural protocol learning for a MAC scenario as in~\cite{Hoydis_2021_arxiv}. In Sec. \ref{subsec:scenario}, we explain the scenario for MAC and the KPIs for evaluating the protocols. In Sec. \ref{subsec:NPM}, we train an NPM by using a multi-agent NN in an MARL environment. In Sec. \ref{subsec:NPM_Limit}, we explain the potentials and limitations of an NPM.

\subsection{A Two-User MAC Scenario}\label{subsec:scenario}

We consider a cellular BS serving two UEs that are contending for the same frequency band at a communication cycle $t=0,1,2,...,T$. A single unit of data that UE $i$ can send within one communication cycle is termed its service data unit (SDU), and it arrives at a rate of $\lambda_i$ (SDU per communication cycle) right before the beginning of a communication cycle, until a total of $D_{\text{max}}$ SDUs arrive. To temporarily store the SDUs, UE $i$ has a buffer with capacity $b_{\max}$. The buffer starts empty, and when an SDU arrives, it occupies one buffer slot; and when the SDU is sent, the corresponding buffer slot is freed. CMs are exchanged to coordinate the UEs to send their SDUs without collision, which occurs when they try to send their SDUs at the same cycle. We assume that the CMs are collision-free and error-free, by using low modulation level and advanced linear block error correcting schemes, e.g.~\cite{PolarCode5G_2021}, but SDUs can experience a block error, which occurs at the rate of $\epsilon$. As illustrated in Fig. \ref{Figure_Cycle}, during a communication cycle, both the control plane and user plane communication take place between each UE and the BS. A communication cycle consists of the following 4 phases: 
\begin{enumerate}[label=(\roman*)]
    \item \textbf{UCM} from UE to BS: UEs communicate via UCMs before sending an SDU to inform the BS about their current buffer state.
    \item \textbf{DCM} from BS to UE: the BS sends DCMs to the UEs to coordinate them in their decisions of sending the SDU or not. 
    \item \textbf{Actions}: UEs either stay silent, access the channel, or discard an SDU. First, nothing happens when a UE stays silent. Second, an SDU in the buffer is sent to the BS in FIFO manner, when the UE decides to access. Third, the newly incoming SDU is discarded when the buffer is already full.
    \item \textbf{ACK/NACK} signals from BS to UE: An ACK signal is sent to notify the UE that its SDU has been received without collision and decoded without error. A NACK signal is sent when the SDUs are lost due to collision, or due to a block error.
\end{enumerate}
Meanwhile, because a cycle is shared between the control and user plane data, the communication efficiency of the CMs is directly related to the amount of data that can be packed into an SDU. Consequently, the main KPI is the goodput $n_R/T$, where $n_R$ is the number of successfully received SDUs. The system objective is thus to maximize the goodput under the efficient usage of communication, memory, and computation resources.

\subsection{NPM Construction and Operation via Multi-Agent Reinforcement Learning}\label{subsec:NPM}

We ground the MAC system model into an MARL environment as in~\cite{Hoydis_2021_arxiv}, referred to as neural protocol learning. RL is instrumental in modeling the behaviors of agents in an environment by rewarding desired behaviors and penalizing undesired ones. A multi-agent DQN is chosen in particular to construct an NPM, because it leverages NNs and experience replay to approximate the Q-function in an environment grounded manner; and the problem is formulated as a decentralized partially observable Markov decision process (Dec-POMDP) with a communication channel between the UEs and the BS~\cite{DecPOMDP_2008}. Consequently, NPM follows centralized training with decentralized execution (CTDE) to address non-stationarity~\cite{Hoydis_2021_arxiv}.

\subsubsection{States, Actions, Observations, and Rewards}
UEs and BS are decentralized agents indexed by $i \in \{1,2,3\}$, $i=3$ being the BS. Agents make partial observations on their current state by observing the input buffer states:
\begin{align}
    \bb =[b_1,\;b_2] 
\end{align}
where $b_i=0,1,2,...,b_{\max}, \; \forall i\in\{1,2\}$. The actions determine what the UE does to its SDU. Depending on $\bb$ and the current protocol, the UE can choose its actions as follows:
\begin{align}
    a_i=
    \begin{cases}
       S,  \quad$silence$,\\
       A,  \quad$access the channel and send the oldest SDU from the buffer$,\\
       D,  \quad$discard the oldest SDU from the buffer$.\\
    \end{cases}
\end{align}
For notational clarity, we denote the action where UE $i$ chooses $S$, $A$, and $D$, as $a_i^S$, $a_i^A$, and $a_i^D$, respectively. When UE $i$ selects $a_i^A$, the buffer state $b_i$ is decremented by $1$, unless a new SDU arrives before the next cycle\footnote{In this work, techniques are targeted toward resource-constrained Internet-of-Things devices, and it is implicitly assumed that no retransmission is allowed to avoid excessive resource consumption~\cite{IoT_Retransmit}. It is interesting to extend the current design towards incorporating retransmission schemes, e.g., random backoff and automatic repeat request, which is outside the scope of the current work.}. According to the contention and decoding result, the following observation is available at the BS:
\begin{align}
    o=
    \begin{cases}
       $idle,  no SDU is received$,\\
       $ACK$i$,  received and decoded SDU from UE $1,\\
       $NACK,  failed to receive or decode an SDU$.\\
    \end{cases}
\end{align}
According to the BS' observation and the actions taken by the UEs, a reward is given to the agents by a central critic. The rewards are defined as follows:
\begin{align}
    r = 
    \begin{cases}
        +\rho_1, $ if one SDU packet is successfully received,$\\
        -\rho_2, $ if a UE discards an SDU packet from its buffer,$\\
        -1, $ otherwise.$
    \end{cases}
\end{align}

\subsubsection{NPM Learning} NPM aims to maximize the average system reward by approximating the optimal Q-function for a given environment. The Q-function is the expectation of the sum of discounted future reward~\cite{sutton2018reinforcement}. To construct an NPM for such a scenario, we optimize the DQN weight parameters $\bw$ illustrated in Fig. \ref{Fig_Schematics}(a) in a centralized manner, with a central critic calculating the reward. The NPM architecture consists of two NN segments for each UE, and one NN segment for the BS. Each NN segment consists of two hidden layers, and the output layer pertaining to a CM or the Q-values. The layers are activated with ReLU, and the hyperparameters are tuned to maximize the average goodput. For the NN segments, $16$ nodes are used for each hidden layer and $8$ nodes are used for the outputs. Detailed hyperparameter settings are explained in Sec. \ref{Sec:Disc_Experim}.

\subsubsection{NPM Operation} The CMs and actions for UE $i\in\{1,2\}$ are chosen according to the following functions, which correspond to the last activation function of each NN segment: 
\begin{align}\label{eq:NPMGU}
    \text{(NPM UCM)\quad} &u_i = g^U_i(b_i|\bw^U), \\ \label{eq:NPMGD}
    \text{(NPM DCM)\quad} &d_i = g^D_i(u_1, u_2|\bw^D), \\\label{eq:NPMGA}
    \text{(NPM Action)\quad} &a_i = g^A_i(d_i|\bw^A),
\end{align}
where $\bw^U$, $\bw^D$, $\bw^A$ are the NN segment parameters within the NPM. As shown in Fig. \ref{Fig_Schematics}(a), the following NPM operation completes a communication cycle:
\begin{enumerate}[label=(\roman*)]
    \item \textbf{UCMs}: the upper NN segment of each UE $g^U_i$ transforms its buffer state $b_i$ into a vector of activation values $u_i$, where each activation value is a positive decimal number represented by a floating point $32$ (FP$32$) data type. The BS concatenates the activations from the UEs into $\bu = [u_1,\;u_2]$.
    \item \textbf{DCMs}: the BS uses its NNs $g_1^D$ and $g_2^D$ to transform $\bu$ into another activation vector $\bd = [d_1,\;d_2]$. UE $1$ gets $d_1$ from the BS, and UE $2$ gets $d_2$.
    \item \textbf{Actions}: each UE uses its bottom NN segment $g^A_i$ to transform the vector into three activations that correspond to the Q-value of each action that the UE could take, and decide the action $a_i$ with the highest Q-value, as its action for this cycle.
\end{enumerate}

\begin{figure}[t]
  \centering
  \includegraphics[width=15cm]{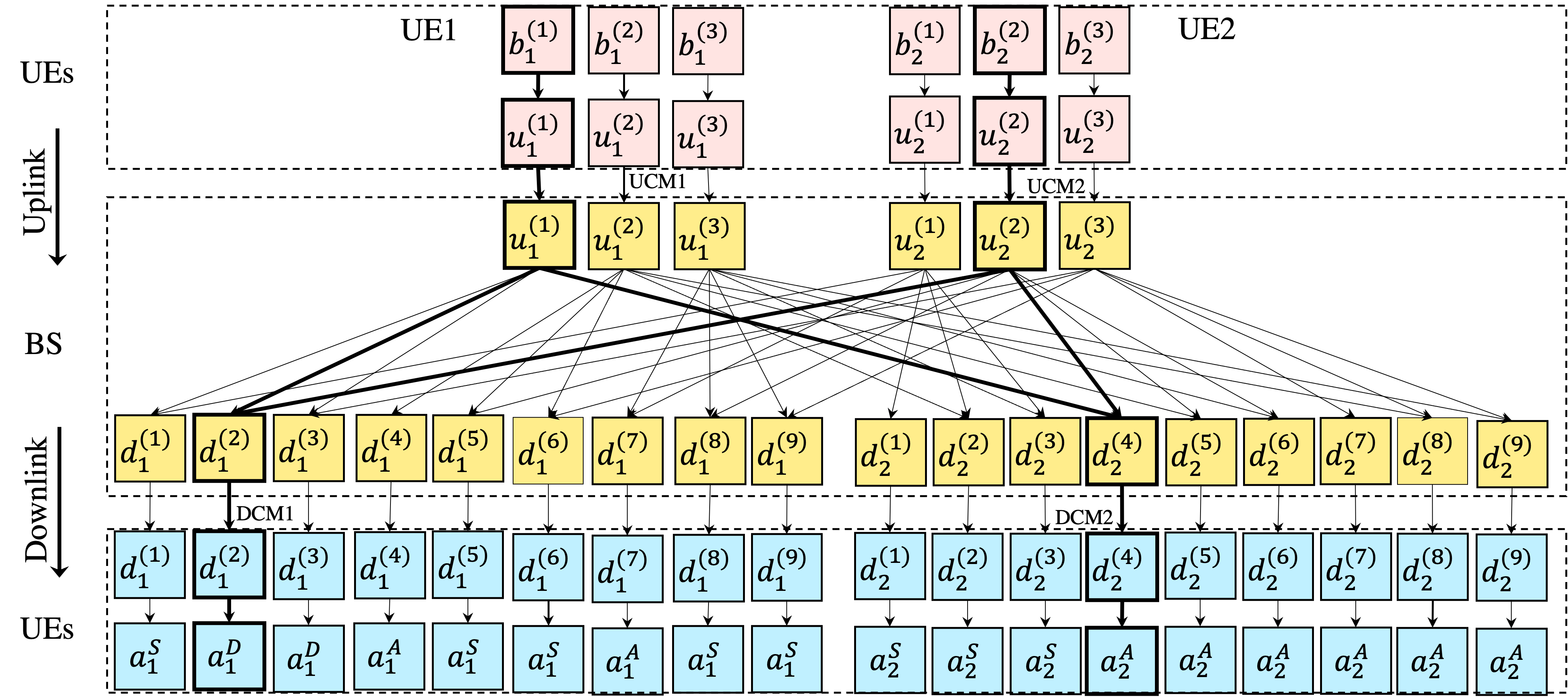}
  \caption{Vocabularies and connections extracted from an NPM when $N=2$ and $|\vB_1|=|\vB_2|=3$. The bold arrows and boxes indicate the vocabularies being activated with the current input $b_1^{(1)}$ and $b_2^{(2)}$. The inference result makes UE $2$ to remain silent whereas UE $1$ accesses the channel.} \label{Fig_ProLog_SPM}
\end{figure}

\subsubsection{Vocabulary and Connection Extraction}
As shown in Fig. \ref{Fig_Schematics}, an NPM could be used for extracting vocabularies and their connections by treating it as a simulator. Then, as illustrated in Fig. \ref{Fig_ProLog_SPM}, we can obtain an NPM extract by logically connecting all activations that correspond to input buffer states, UCMs, DCMs, and actions that are inferred simultaneously during the episodic memory, which is implemented to store the historical data on the states, intermediate activations, and actions. For a buffer state denoted by $b_i^{(k)}$, which is the $k$-th buffer state among $K$ buffer states of UE $i$, a distinct UCM denoted by $u_i^{(k)}$ is inferred by \eqref{eq:NPMGU}. Next, for each distinct UCM pair $\bu^{(k)}$, a distinct DCM $d_i^{(k)}$ is inferred by \eqref{eq:NPMGD}. Lastly, an action $a_i$ is inferred from the DCM. To cast these directed relationships as a graph, we connect the vocabularies that occur simultaneously and consequentially with a single-headed arrow. For example, if the buffer states are $b_1^{(1)}$ and $b_2^{(2)}$, then $u_1^{(1)}$, $d_1^{(2)}$, $a_1^{D}$ are consequentially inferred for UE $1$, and $u_2^{(2)}$, $d_2^{(4)}$, $a_2^{A}$ are inferred for UE $2$. Graphing these sequential inference relationships for every input buffer state gives the NPM extract expressed as a symbolic graph in Fig. \ref{Fig_ProLog_SPM}.

\subsection{Potentials and Limitations of NPMs} \label{subsec:NPM_Limit}

\begin{figure}[t]
  \centering  
  \includegraphics[height=5.5cm]{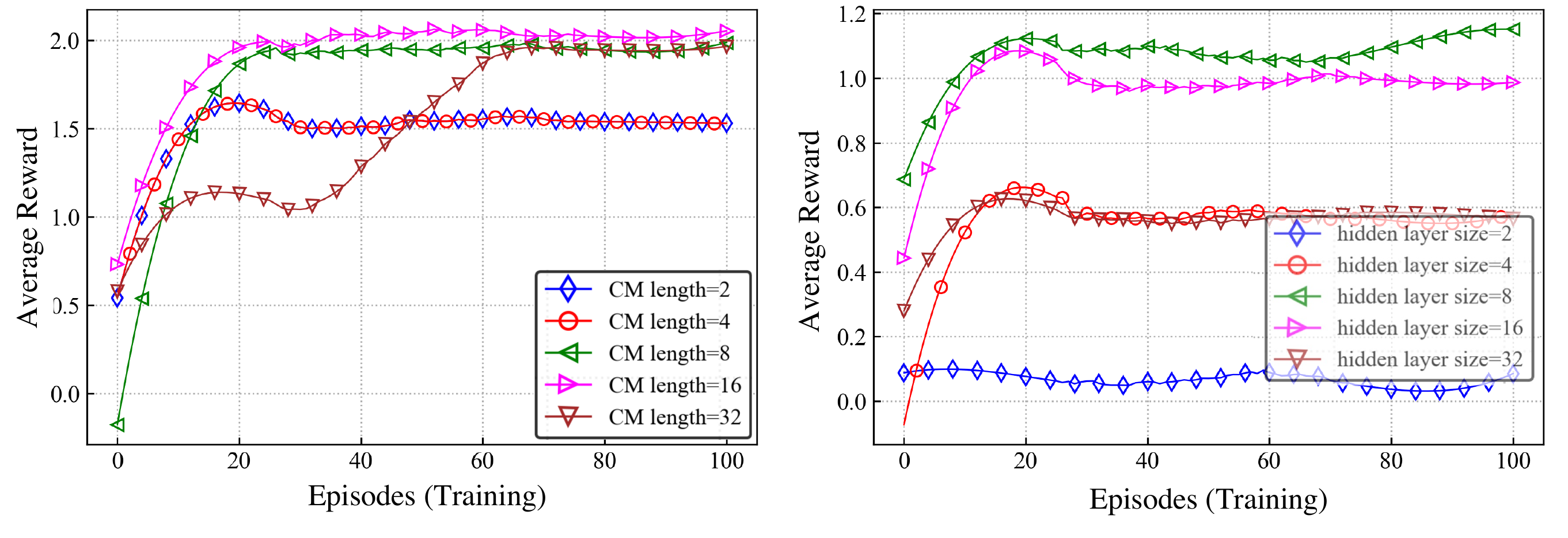}
  \caption{NPM needs CM length of at least 8 output activation nodes (left), and 16 intermediate activation nodes (right) to ensure good training results on average.}   \label{Fig_NPM_Limit}
\end{figure} 

Via overparametrization, NPMs can be trained to achieve high performance for the environment they are grounded in, e.g.~\cite{Hoydis_2021_arxiv, Han_2020_commag, Liu_2006_RLMAC, lee_2021_LEO}. However, overparametrization results in the construction of inefficient NPMs, which exerts excessive communication, computation, and memory overhead during each communication cycle. The overparametrized layers and nodes directly translate into increased storage and computation load, which makes the system less energy efficient and hinders low latency communication. Additionally, as shown in Fig. \ref{Figure_Cycle}, increased communication overhead due to longer CM is critical because it reduces the time for an SDU to be sent, resulting in a degraded throughput. Fig. \ref{Fig_NPM_Limit}(a) shows the average utility when training with different numbers of activation nodes. In particular, the number of activation nodes at the intermediate output layers directly translates into the length of CMs. Empirically, we observe that DQNs need at least $8$ activations to achieve the average reward of $2$. This means that an NPM has the minimum control length requirement of $32$ Bytes, which becomes a huge overhead to send in every communication cycle.

Even if we use the symbolic graph version instead of a full NPM, an NPM encounters communication and memory inefficiencies for larger buffer capacity and more users. In fact, the UCM vocabulary size increases combinatorially and DCM vocabulary size increases exponentially because a DCM is required for every combination of UCM from the UEs. Consequently, the control plane complexity is increased due to the number of connections needed to explain the relationship between all UCMs and DCMs. The communication inefficiency due to larger vocabulary size, and increased control plane complexity due to the complex protocol model are incompatible for 6G, which is projected to have a unified control plane architecture~\cite{6G_Requirements_2020} with distributed core units capable of computing at sub-$1$ms control plane latency~\cite{6G_Control_Plane_2020, 6G_ML_Challenges_2020}. Thus, it is necessary to devise methods where we can emulate the performance of an NPM while ensuring communication and memory efficiency.

\section{SPM: Semantic Protocol Model via \\ Probabilistic Logic Programming Language} \label{Sec:S2PM}


To obviate the shortcomings of an NPM, we propose the construction of an SPM. We make SPM communication and memory efficient by merging UCM and DCM vocabularies that exhibit redundant semantics. To greatly reduce the vocabulary sizes, we combine two merging techniques, which let us benefit from the reduction capability of both merging techniques. However, combining the two leads to problematic situations where signaling vocabularies become polysemous, which we refer to as the \textit{polysemy problem}. We resolve such problem by formulating SPM rules based on probabilistic logic clauses and utilizing logical inference based on contextual information stored in the empirically derived conditional probabilities.

\subsection{An Overview of ProbLog and SPM Construction Procedure}
We deconstruct an NPM into an SPM by extracting the logical relationships from the episodic memory of MARL. To express the logical aspects of an SPM such as vocabularies, clauses, predicates, and rules, and to describe an SPM's construction procedure succinctly and effectively, we adhere to the syntax of ProbLog~\cite{problog} as follows:
\begin{enumerate}[label=(\roman*)]
    \item \textbf{\emph{Vocabularies} for CMs}: We use four types of vocabularies denoted by $b_i$, $u_i$, $d_i$, $a_i$, for input buffer states, UCMs, DCMs, and actions of UE $i$, respectively. Given two vocabularies that are connected as a causal relationship, i.e. ``$\to$'', a vocabulary is referred to as the $\mathsf{Tail}$ if it is the \textit{cause} of a causal relationship, and referred to as $\mathsf{Head}$ if it is the \textit{effect}. The vocabularies are extracted in Sec. \ref{subsub:vocab_ext} and merged in Sec. \ref{subsub:act_merge}-\ref{subsub:conn_merge}.
    
    \item\textbf{\emph{Clauses} for CM Relations}
    A probabilistic logic clause $c$ that connects a $\mathsf{Head}$ and a $\mathsf{Tail}$, i.e. ``with probability $p$, $\mathsf{Head}$ is true, if $\mathsf{Tail}$ is true'' is expressed as $c = \langle p:: \mathsf{Head}\text{ :- } \mathsf{Tail} \rangle$ in ProbLog. Each vocabulary in $c$ could be accessed by $c^\mathsf{H}=\mathsf{Head}$ and $c^\mathsf{T}=\mathsf{Tail}$, and the probability is accessed by $c^\mathsf{P}=p$, where $\mathsf{P}$ denotes the clause's probability of being true, i.e. the \textit{truth probability}. The semantic clauses are defined in Sec. \ref{subsub:clause}.
    
    \item \textbf{\emph{Predicates} for Semantic Clause Clustering}: We use four predicates $\mathsf{is Input}(\cdot)$, $\mathsf{isUCM}(\cdot)$, $\mathsf{is DCM}(\cdot)$, and $\mathsf{is Action}(\cdot)$ that are used on the vocabularies to describe its type. An uplink clause $\a$ is described by $\mathsf{isInput}(\a^\mathsf{T}) = \mathsf{isUCM}(\a^\mathsf{H}) = \mathsf{true}$, a downlink clause $\b$ is described by $\mathsf{isUCM}(\b^\mathsf{T}) = \mathsf{isDCM}(\b^\mathsf{H}) = \mathsf{true}$, and an action clause $\c$ is described by $\mathsf{isDCM}(\b^\mathsf{T}) = \mathsf{isAction}(\b^\mathsf{H}) = \mathsf{true}$. Subsequently, we can cluster the clauses according to the clause types $\a$, $\b$, and $\c$. The predicates are used for clustering clauses in Sec. \ref{subsub:clause}.
    
    \item \textbf{\emph{Rules} and \emph{Entailment} for SPM Construction}
    : We define a rule $\mathcal{R}(\cdot)$ as a sequence of simultaneously occurring clauses that are logically consequential from an input. For example, given a simple protocol that consists of the two clauses $\langle \mathsf{A} \text{ :- } \mathsf{B}\rangle$ and $\langle \mathsf{B} \text{ :- } \mathsf{C}\rangle$, we could consequentially derive that the two clauses are true if we know that $\mathsf{A}$ is true. We describe this relationship as \textit{entailment}, denoted by $\mathcal{R}(\mathsf{A}) \vDash \langle \mathsf{A} \text{ :- } \mathsf{B}\rangle, \langle \mathsf{B} \text{ :- } \mathsf{C}\rangle$, where $\vDash$ means to logically entail. SPM rules are formulated in Sec. \ref{subsub:rule}.
    
    
\end{enumerate}
The SPM construction requires the following procedure, which is expressed with ProbLog for clarity. First, we extract the vocabularies from the episodic memory of MARL. Second, we merge the vocabularies that exhibit redundant semantics according to two schemes. Third, we define the three logic clause types $\a$, $\b$, and $\c$ that we can use for clustering the logical clauses according to the operation it entails such as uplink, downlink, and action. Fourth, we formulate SPM rules $\R(\cdot)$ that act as the smallest element of SPM operation. Lastly, an SPM $\SPM$ is constructed as the set of all rules that are formulated from each input buffer state.

\subsection{SPM Construction}
\label{subsec:S2PM_Seq_Clause}

\begin{figure*}[t]
  \centering  
  \includegraphics[height=5cm]{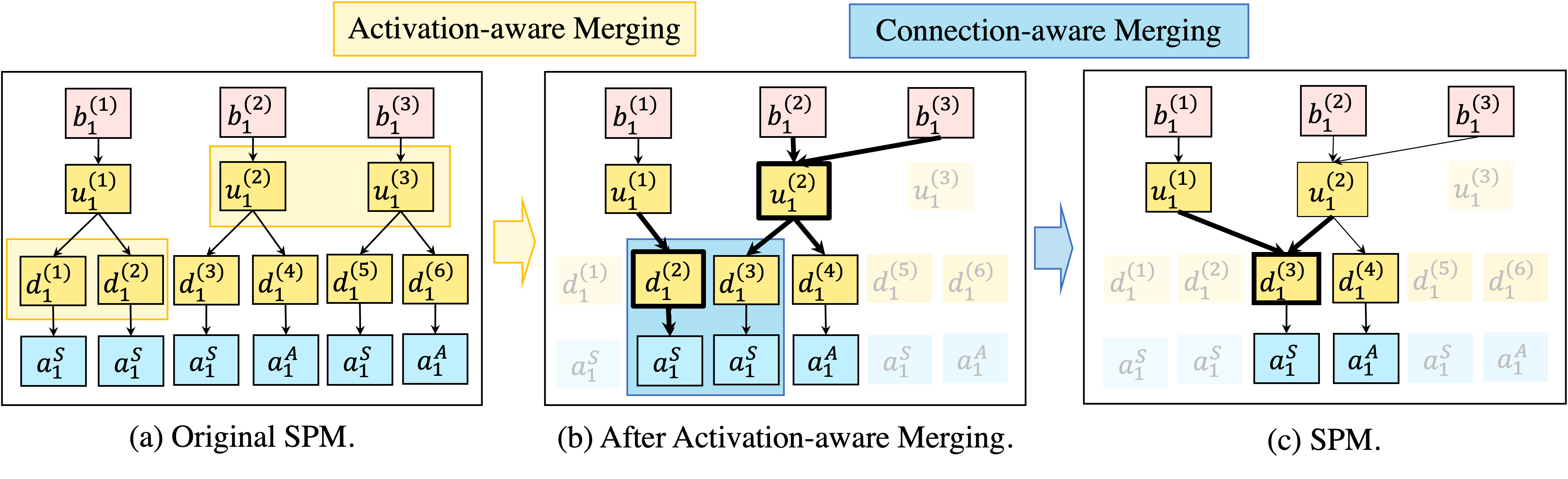}
  \caption{Transformation process: blurred vocabulary are merged; (a) the original SPM is (b) merged according to their activation patterns, and (c) merged according to their vocabulary connections.} \label{Fig_SPM_transformation}
\end{figure*}

Fig. \ref{Fig_SPM_transformation} illustrates the SPM merging process needed for constructing a communication and memory efficient SPM. The main idea is to merge the UCM and DCM vocabularies if they exhibit semantic redundancy in terms of activations and connections. This merging process results in fewer UCM and DCM bits to be transmitted for each communication cycle, and a reduction of clauses required for expressing a model. Despite its efficiency, merging CMs, which contain inter-UE semantics, introduces the cases where the DCMs become polysemous. We resolve this issue by exploiting contextual information embedded in the empirical distributions of UCM and DCM realizations.

\subsubsection{Vocabulary Extraction}\label{subsub:vocab_ext}

We define the input state, UCM, DCM, and action vocabularies as possible outputs from the NPM generators \eqref{eq:NPMGU}-\eqref{eq:NPMGA}, i.e. $b_i \in \vB_i$, $u_i=g^U_i(b_i) \in \vU_i$, $d_i=g^D_i(\bu) \in \vD_i$, and $i \in \{1,2\}$ and $a_i=g^A_i(d_i) \in \vA_i$ for $i\in \{1,2\}$. First of all, the vocabularies we need to consider can be restricted to those extracted from the episodic memories within the experience replay buffer. To elaborate, assuming that the memory of the experience replay is available, we can consider the input states experienced in the memory, i.e. $\vB_1$ and $\vB_2$, to be the feasible domain that is plugged into \eqref{eq:NPMGU}. This is because the other states are very unlikely to be traversed. Even when the replay memory is unavailable, we can obtain one by running a few test trials. The rest of the procedure merges these vocabularies according to their semantics, and embeds them into SPM clauses that contain the semantic information regarding their relationships.

\subsubsection{Activation-Aware Vocabulary Merging}\label{subsub:act_merge}
As Fig. \ref{Fig_SPM_transformation}(c) illustrates, we merge the CM vocabularies by using their activation pattern, i.e. the location of non-zero elements in the activation vector extracted by $\tv_i=\tH(v_i)$, where we assume ReLU activations, and $\tH(v_i)$ is the Heaviside step function $\tH(x) = \ind_{\{x>0\}}$ applied to each decimal number within the vocabulary $v_i$. For example, suppose that $v_1$ is the vocabulary expressed with $8$ decimal numbers:
$[0.382,4.292, 0, 0, 1.249, 0, 0, 0]$. The activation pattern of $v_1$ is the vocabulary $[1, 1, 0, 0, 1, 0, 0, 0]$. Without activation and parameter quantization techniques~\cite{Deep_Compress_2016_ICLR, Gauss_Quantization_2017_CVPR}, the performance of NPMs depends heavily on the precision of the activation signals. Nevertheless, from empirical observations, we hypothesize that there is the possibility of reducing the vocabularies greatly by exploiting the activation patterns. Consequently, we merge the UCM and DCM vocabularies denoted by $u^{(k_1)}_i, u^{(k_2)}_i$ and $d^{(\l_1)}_i, d^{(\l_2)}_i$, respectively, for all UCMs index by $ k_1$ and $k_2$, and DCMs indexed by $\l_1$ and $\l_2$, if they have the same activation pattern as follows:
\begin{align}
    \text{(Activation-aware UCM merging)\quad} &u^{(k')}_i \xleftarrow{}  u^{(k_1)}_i, u^{(k_2)}_i, \;\; \text{if} \;\;\tH(u^{(k_1)}_i)=\tH(u^{(k_2)}_i) ,\label{Heavy_UCM}\\
    \text{(Activation-aware DCM merging)\quad} &d^{(\l')}_i \xleftarrow{} d^{(\l_1)}_i, d^{(\l_2)}_i, \;\; \text{if} \;\;\tH(d^{(\l_1)}_i)=\tH(d^{(\l_2)}_i),\label{Heavy_DCM}
\end{align}
By using fewer vocabulary for UCM and DCM messages, the logical rules become overlapped with each other, as shown in Fig. \ref{Fig_SPM_transformation}(b). Because of the other UE's impact on the decision of a DCM, the merging process makes the choice of some actions arbitrary due to polysemous DCMs, in contrast to the deterministic choices made by the NPM. To disambiguate the polysemous vocabularies, we devise a method to determine which CMs and actions have more contextual meaning for the current operation, by exploiting their likelihood represented by the clauses' conditional probabilities.

\subsubsection{Connection-Aware Merging} \label{subsub:conn_merge}
To further increase communication efficiency, we merge the DCMs, UCMs, and clauses altogether, identifying cases when the connected vocabularies are identical, as illustrated in step (b) of Fig. \ref{Fig_SPM_transformation}. Firstly, the DCM vocabularies denoted by $\td^{(\l_1)}_i$ and $\td^{(\l_2)}_i$ are merged when the set of action vocabularies they are connected to are identical. The following update rule is applied for all DCMs indexed by $\l_1$ and $\l_2$ to merge the DCM vocabulary and its connections:
\begin{align}
    \text{(Connection-aware DCM merging)\quad}&\td^{(\l')}_i \xleftarrow{} \td^{(\l_1)}_i, \td^{(\l_2)}_i, \;\; \text{if} \; \;\vA^{(\l_1)}_i=\vA^{(\l_2)}_i,\label{eq:Twin_alpha}
\end{align}
where $\vA^{(\l)}_i= \{a_i \in \vA_i | \c_i^\mathsf{T} =  \td^{(\l)}_i,  \c_i^\mathsf{H} = a_i \}$. Secondly, the UCM vocabularies denoted by $\tu^{(k_1)}_i$ and $\tu^{(k_2)}_i$ are merged when the set of DCM vocabularies they are connected to are identical. The following update rule is applied for all UCMs indexed by $k_1$ and $k_2$ to merge the UCM vocabulary and its connections:
\begin{align}
    \text{(Connection-aware UCM merging)\quad}&\tu^{(k')}_i \xleftarrow{}  \tu^{(k_1)}_i, \tu^{(k_2)}_i, \;\; \text{if} \;\; \vD^{(k_1)}_i=\vD^{(k_2)}_i, \label{eq:Twin_beta_DCM}
\end{align} where $\vD^{(k)}_i= \{\td_i \in \vD_i | \b_i^\mathsf{T} =  \tu^{(k)}_i,  \b_i^\mathsf{H} = \td_i \}$. Applying connection-aware merging to UCMs reduces the UCM vocabulary further, but it makes the UCMs polysemous. Again, to resolve the polysemy problem, we embed additional semantics within the downlink clauses via the conditional probability and utilize them with the SPM rule and operation.

\subsubsection{Clause Definition and Semantic Clustering}\label{subsub:clause} To construct an SPM clause, we need to decide the type of clauses we will construct, dictating the logical relationships of interest as follows. Technically, we could construct a clause that explains the relationship between any activation layer to any other activation layer, but we only focus on the layers corresponding to the vocabularies extracted above. We define the clause sets $A_i,B_{i,j},\Gamma_i \; \forall i,j \in \{1,2\}$ to describe the logical relationships of interest. The uplink clause set $A_i$ contains clause $\a_i$, named \textit{uplink clause}, which connects UE $i$'s input buffer state $b_i$ to its UCM $u_i$. The downlink clause set $B_i$ contains clause $\b_i$, named \textit{downlink clause}, which connects the UCM $u_j$ from UE $j \in \{1,2\}$ to the DCM $d_i$ that will be sent to UE $i \in \{1,2\}$. The action clause set $\Gamma_i$ contains clause $\c_i$, named \textit{action clause}, which connects UE $i$'s DCM $d_i$ to its action $a_i$.

To resolve the polysemy problem, we construct SPM clauses that express the level of truth in a logical connection. We use ProbLog to express the uplink, downlink, and action clauses as follows:
\begin{align} \label{eq:b_u_Prob}
    \text{ (Uplink clause)\quad} &\text{\colorbox{MistyRose1}{$\ha_i = \langle 1\text{ :: }\tu_i\text{ :- } b_i\rangle$}}, \\ \label{eq:u_d_Prob}
    \text{ (Downlink clause)\quad} &\text{\colorbox{LightGoldenrod1}{$\hb_{i,j}= \langle \mathsf{Pr}(\td_i|\tu_j)\text{ :: }\td_i\text{ :- } \tu_j\rangle$}}, \\ \label{eq:d_a_Prob}
    \text{ (Action clause)\quad}&\text{\colorbox{LightBlue1}{$\hc_i = \langle \mathsf{Pr}(a_i|\td_i)\text{ :: }a_i\text{ :- }\td_i\rangle$}},
\end{align}
where $\a_i \in A_i$, $\b_{i,j} \in B_{i,j}$, $\c_i \in \Gamma_i$, $u_i\in \mathcal{U}_i$, $d_i \in \mathcal{D}_i$, $a_i \in \mathcal{A}_i$, and $i,j \in \{1,2\}$, and the conditional probabilities are calculated empirically by simulating over the domain given by the episodic memory. The probability is $1$ for \eqref{eq:b_u_Prob} because $b_i$ and $u_i$ are always selected simultaneously; the conditional probability of \eqref{eq:u_d_Prob} is empirically calculated as the following ratio: the number of times $u_j$ and $d_i$ are simultaneously selected divided by the total number of times $u_j$ is selected; lastly, the conditional probability of \eqref{eq:d_a_Prob} is calculated as the following ratio: the number of times $d_i$ and $a_i$ are simultaneously selected divided by the total number of times $d_i$ is selected. 



\begin{figure}[t]
  \centering
  \includegraphics[height=3cm]{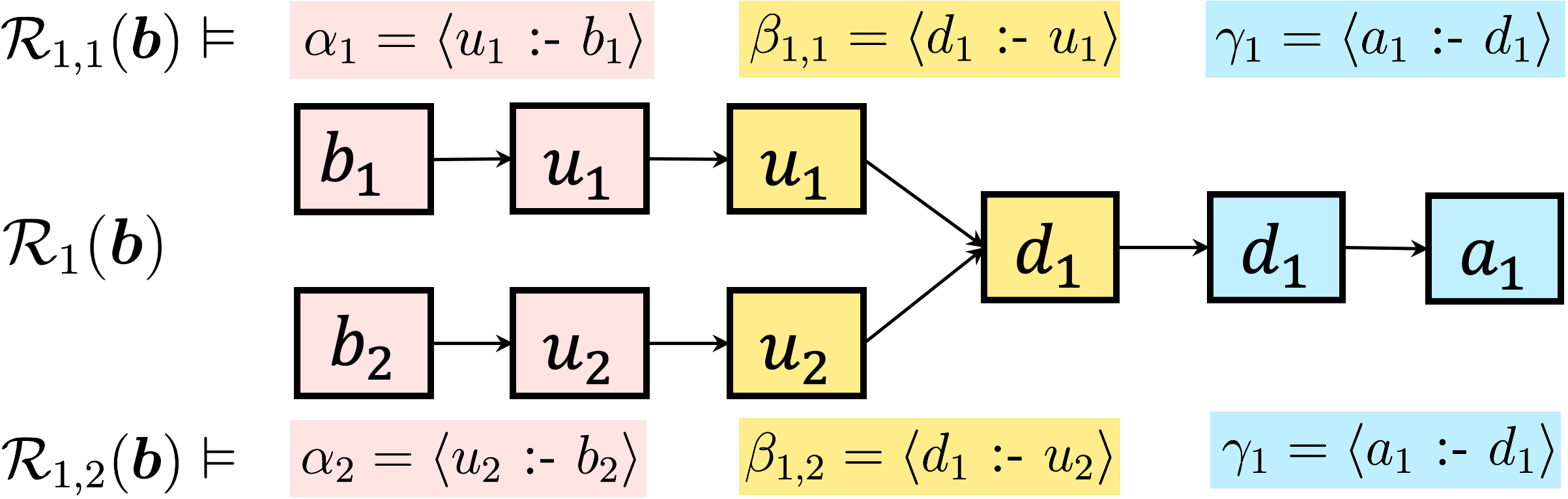}
  \caption{An example of rules that entail the sequential logical connections of $\a$, $\b$, and $\c$. The clauses in the example are deterministic for simplicity.} \label{Fig_SPM_Rule}
\end{figure}

\begin{figure}[t]
  \centering
  \includegraphics[height=7cm]{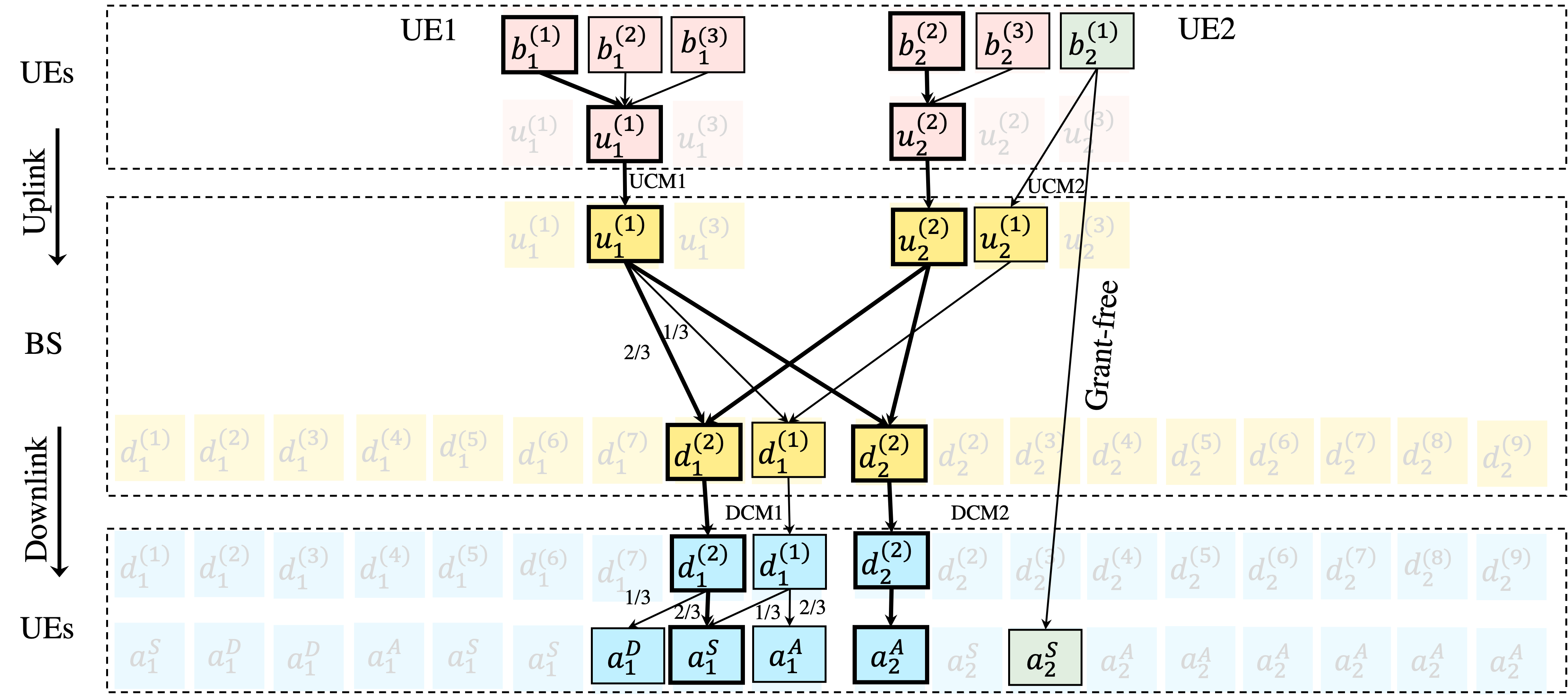}
  \caption{An example of SPM when $N=2$ and $B=3$, which is the transformed version of Fig. \ref{Fig_ProLog_SPM}. The bold arrows and boxes indicate the decisions made using the maximum truth probability for input $b_1^{(1)}$ and $b_2^{(2)}$. The operation result makes UE $1$ remain silent whereas UE $2$ accesses the channel.} \label{Fig_S2PM}
\end{figure}

\subsubsection{SPM Rule Formulation and SPM Construction}\label{subsub:rule}
We construct an SPM rule, which is the smallest building block of an SPM, that entails a logical sequence of clauses that connects the UCM, DCM, and action vocabularies that are causal from an input buffer state, as illustrated in Fig. \ref{Fig_SPM_Rule}. Specifically, plugging in a particular $b_i$ to the NPM, we can obtain the selected vocabularies that correspond to the heads and tails of $\a$, $\b$, and $\c$ that occur simultaneously. To express this consequential relationship, we define an SPM rule for the action of UE $i$ that originates from UE $j$'s state as the following set-valued function that entails the clauses as follows:
\begin{align} \label{eq:StoPRule}
    \mathcal{R}_{i,j}(\bb) \vDash \ha_j, \hb^{(1)}_{i,j},...,\hb^{(L)}_{i,j}, \hc^{(1,1)}_i,...,\hc^{(L,M)}_i 
\end{align}
where the conditions for $\ha_j,\hb_{i,j}^{(\l)},\hc^{(\l,m)}_i$ entailed by $\mathcal{R}_{i,j}(\bb)$ must be met for all $\l=1,2,...,L$ and $m=1,...,M$ as follows:
\begin{align}\label{eq:SPM_rule_condition1}
    &\text{\colorbox{MistyRose1}{$\ha_j^\mathsf{T}$}} = b_j, \\ \label{eq:SPM_rule_condition2}
    &\text{\colorbox{MistyRose1}{$\ha_j^\mathsf{H}$}} = \text{\colorbox{LightGoldenrod1}{${\hb_{i,j}^{(\l)}}^\mathsf{T}$}}\ = \tu_j,\; \forall \l=1,2,...,L \\\label{eq:SPM_rule_condition3}
    &\qquad\quad \text{\colorbox{LightGoldenrod1}{${\hb_{i,j}^{(\l)}}^\mathsf{H}$}}= \text{\colorbox{LightBlue1}{${\hc_i^{(\l,m)}}^\mathsf{T}$}}=d_i,\; \forall m=1,...,M,
    \\\label{eq:SPM_rule_condition4}
    &\qquad\qquad\qquad\quad\text{\colorbox{LightBlue1}{${\hc_i^{(\l,m)}}^\mathsf{H}$}} = a_i^{(m)}, 
\end{align}
where ${\hb_{i,j}^{(\l)}}^\mathsf{P} > 0$, ${\hc_i^{(\l,m)}}^\mathsf{P} > 0$, and the vocabulary belonging to the same clause is highlighted with the same color. The conditions \eqref{eq:SPM_rule_condition1}-\eqref{eq:SPM_rule_condition4} describe the consequential relationship between the clauses. In \eqref{eq:SPM_rule_condition1}, the state $\bb$ first entails a clause $\a_j$ that has $\bb$ as its tail; in \eqref{eq:SPM_rule_condition2}, the UCM $u_j$ of the clause $\a_j$, entails all $\b_{i,j}^{(\l)}$ that has UCM $u_j$ as its tail; in \eqref{eq:SPM_rule_condition3}, the DCM of each $\b_{i,j}^{(\l)}$ entails all $\c_i^{(\l,m)}$ that has the same DCM as $\b_{i,j}^{(\l)}$; lastly, in \eqref{eq:SPM_rule_condition4}, each $\c_i^{(\l,m)}$ entails an action $a_i^{m}$ that is implicated with a non-zero probability. An SPM is constructed as the set of all rules that originate from all input buffer state experienced in the episodic memory. Let the SPM rules be defined with \eqref{eq:StoPRule}, the SPM is constructed as follows:
\begin{align}
    \SPM = \bigcup_{i, j, \bb} \mathcal{R}_{i, j}(\bb),
\end{align}
where $\bb \in \vB_1 \times \vB_2$ and $i, j\in\{1,2\}$. 


\subsection{SPM Design Motivation and Empirical Justification}

\begin{figure}[t]
  \centering
  \includegraphics[height=7.4cm]{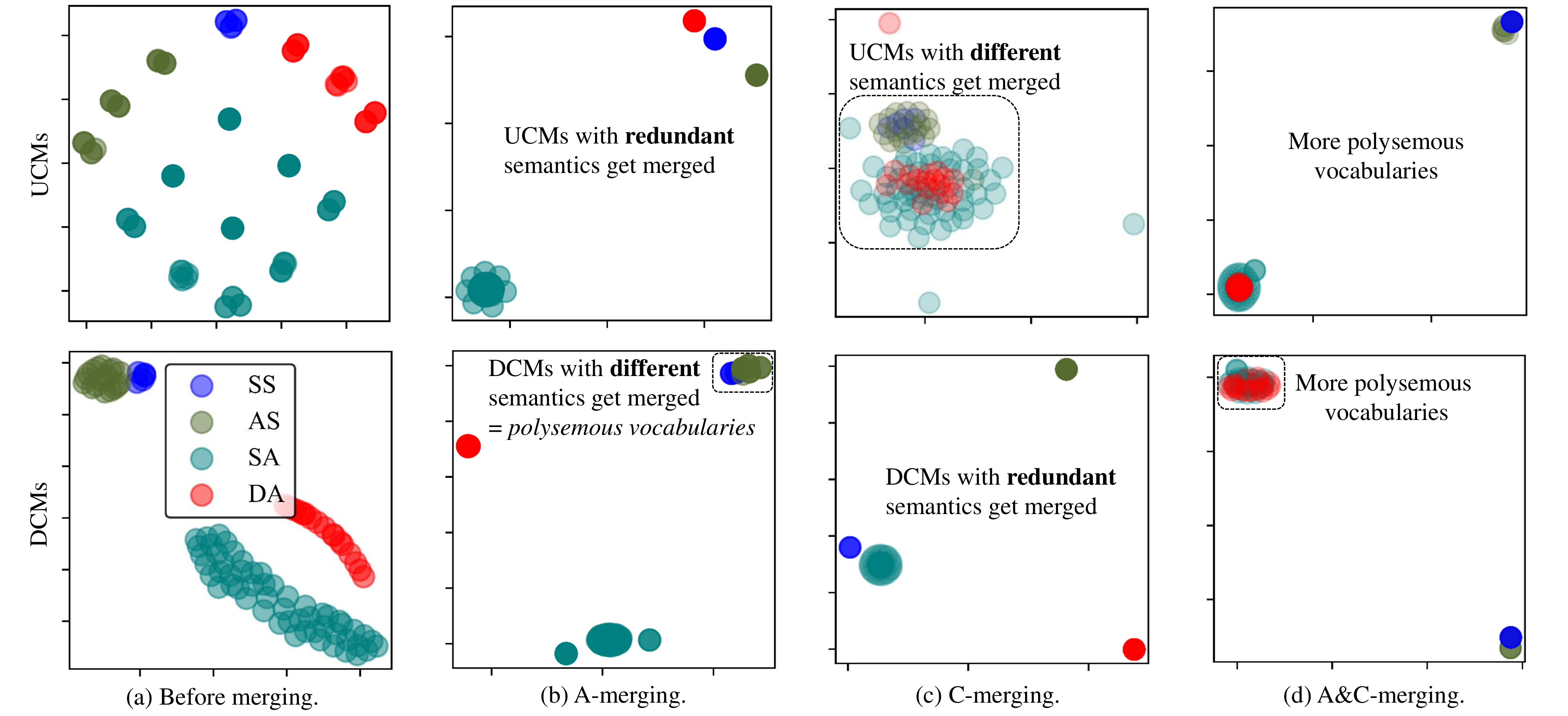} 
  \caption{The t-distributed stochastic neighbor embedding (t-SNE) of concatenated UCM and DCM vocabularies (a) before merging, (b) after activation-aware merging (A-merging), (c) after connection-aware merging (C-merging), and (d) after A-merging and C-merging. The colors indicate the semantics dictated as the action pair $[a_1,a_2]$ of the two UEs taken at that state; and the perplexity parameter is set to $15$.}\label{Fig_S1_Embed}
\end{figure}

For communication and memory efficiency, we have constructed the SPM after merging the vocabularies and connections that exhibit redundant semantics. The total reduction of the vocabulary size can be seen by comparing the graph of an SPM and the blurred graph of the original NPM extracted in Fig. \ref{Fig_S2PM}.
To visualize both positive and negative effect of the merging schemes, we use the t-distributed stochastic neighbor embedding (t-SNE)~\cite{TSNE_2008} graphs of the UCMs and DCMs pairs, i.e. $\bu$ and $\bd$, respectively, as illustrated in Fig. \ref{Fig_S1_Embed}. Fig. \ref{Fig_S1_Embed}(a) shows the t-SNE graphs of vocabularies before their merger, which shows that there are multiple UCM and DCM clusters exhibiting identical semantics, expressed by identical actions pair $[a_1,\; a_2]$ and indicated by the same color. After activation-aware merging, we observe that UCM clusters with the same semantics are merged, but some of the DCM clusters with different semantics exhibits semantic polysemy. On the other hand, connection-aware merging shows promising clustering of DCMs, but introduces severe polysemy for UCMs. Lastly, combining both schemes significantly merges the vocabularies further, but the polysemy occurs for both UCMs and DCMs. The results suggest that there is a minimum number of UCM vocabularies and DCM vocabularies required to express the protocol without any polysemy. However, further reduction gain could be achieved if polysemous vocabularies can be disambiguated.

\subsection{SPM Operation}


To disambiguate the polysemous vocabularies, we utilize the conditional probabilities that contain contextual information regarding the environment. To express the level of confidence in the selection of a vocabulary, the truth probability of the CMs and actions for UE $i$ are calculated as follows when the entailment at state $\bb$ is given by $\hR_{i,i}(\bb) \vDash \ha_i, \hb_{i,i}, \hc_i$ and $\hR_{i,j}(\bb) \vDash \ha_j, \hb_{i,j}, \hc_i$:
\begin{align} \label{eq:Prob_UCM}
    &\mathsf{Pr}(\tu_i|\hSPM, \bb) = \ha_i^\mathsf{P}, \\
    &\mathsf{Pr}(\td_i|\hSPM, \bb) = \hb_{i,i}^\mathsf{P} \hb_{i,j}^\mathsf{P},\label{eq:Prob_dcm}\\
    &\mathsf{Pr}(a_i|\hSPM, \bb) = \hc_i^\mathsf{P},
\end{align}
where $i \neq j \in \{1,2\}$. A higher truth probability indicates higher confidence in the selection of a particular vocabulary for the current context $\SPM$ and $\bb$. Note that \eqref{eq:Prob_dcm} is the joint probability of the event that the same DCM is implicated by the UCMs from UE $i$ and $j$; also, the equality for the second equation holds, because we assume that the probabilities $\hb_{i,i}^\mathsf{P}$ and $\hb_{i,j}^\mathsf{P}$ are independent. Inspired by RL principles, which transform the soft Q-values to hard decisions, we select the CMs and actions according to the maximum truth probability as follows:
\begin{align}\label{eq:S2PM_gen_u}
    \text{(SPM UCM)\quad} &\tu_i = \a_i^\mathsf{H}, \\ \label{eq:S2PM_gen_d}
    \text{(SPM DCM)\quad} &\td_i = \argmax_{\td\in\vD_i} \mathsf{Pr}(\td|\hSPM, \bb), \\ \label{eq:S2PM_gen_a}
    \text{(SPM Action)\quad} &a_i =\argmax_{a\in\vA_i} \mathsf{Pr}(a|\hSPM, \bb),
\end{align}
where $i,j \in \{1,2\}$. The advantages of using the vocabulary with the maximum truth probability are twofold: the computation cost from random sampling is saved, and the operation becomes certain, removing any uncertainty, which is an important trait for a protocol.

Notwithstanding the amount of communication efficiency brought by the schemes explained above, it is best if actions could be taken without a grant given by the DCMs. We enable grant-free communication by comparing the SPM rules, which is referred to as \textit{rule-aware grant-free communication}. When the SPM rule $\R_{i,i}(b_i, b_j)$ is unaffected by $b_j$, $j \neq i$, i.e. $\R_{i,i}(b_i, b_j^{(1)}) = \cdots = \R_{i,i}(b_i, b_j^{(|\vB_j|)})$, we skip the DCM and take an action in a grant-free manner according to the fourth clause type defined as follows:
\begin{align} \label{eq:delta_fa}
    \text{ (Grant-free clause)\quad} &\text{\colorbox{Honeydew2}{$\hd_i = \langle 1\text{ :: }  a_i \text{ :- } b_i \rangle$}},
\end{align}
where $a_i = \c_i^\mathsf{H}$. Compared with our approach that enables grant-free communication with semantic vocabularies using the logical relationship, a tabular RL-based approach in~\cite{hoydis_MAC_journal} uses predefined signaling messages to learn to skip the ACK signal when $\epsilon$ is low, and the deep RL-based approach in~\cite{Hoydis_2021_arxiv} learns emergent signaling messages but the DCM cannot be skipped because of the intrinsic NN structure.

Inspired by natural language based on pragmatics such as contextual meaning to disambiguate polysemous vocabularies, we have disambiguated the UCMs and DCMs via probabilistic logic-based inference in \eqref{eq:Prob_UCM}-\eqref{eq:S2PM_gen_a}. The results from Sec. \ref{Sec:Disc_Experim} will corroborate the probabilisitic logic inference's disambiguation capability. As a summary of the efforts in this section, Table \ref{tab:performance} compares the measured KPIs of an SPM compared with an NPM, such as the goodput, vocabulary sizes, memory requirement, and so forth. The reduced CM length and model storage is consequential to the reduction gain achieved by the merging process, and the measurements in Fig. \ref{Fig_Schematics} are empirically derived from an actual NPM and its \stwo transformation. 

\section{Attributes and Potentials of SPM}\label{Sec:Potentials}

Classical protocols are designed to meet the requirements for a general purpose, whereas NPM is trained to perform well in a specific environment. Inspired by both classical protocols and emergent NPM, an SPM exhibits three major potentials that enable its wider usage: reconfigurability, measurability, and compactness. In this section, we explain each potential and provide an example application that exploits each potential.


\subsection{Reconfigurability}

The classical protocols can be reconfigured to address changes in the environment or system requirement, but an NPM should be retrained with the hyperparameters set accordingly to the new constraint. This process could be taxing because hyperparameters need to be reoptimized, and NPMs need many trials to converge to an optimal solution. Inspired by classical protocols, an SPM can be reconfigured to perform well for a new constraint, by manipulating the logical relationships based on vocabularies, clauses, and rules. As a general principle for manipulation, we need to make the least manipulation steps as possible, to maintain the performance of the original SPM. For example, consider that we want to reduce collisions. We can manipulate the action clause $\c_i \in \Gamma_i$ to reconfigure UE $i$'s action to another action. Let $\c_i^{(\l,A)}$ be the action clause that is entailed by $\bb$, which leads to a collision. We can achieve the desired reconfiguration while making the least manipulation to the SPM by updating the SPM as follows:
\begin{align} \label{eq:CA_Manipulation}
    (\hSPM_i \setminus \hc_i) \cup  \langle \hc_i^\mathsf{P} :: a^{S}_i \text{ :- }d_i\rangle,
\end{align}
where $\hR_{i,i}(\bb) \vDash \ha_i, \hb_{i,i}^{(\l)}, \hc_i^{(\l,A)}$; $\hc_i^\mathsf{H}=a^A_i $ and $\hc_i^\mathsf{T}=d_i $, and the reconfiguration makes UE $i$ to stay silent instead of its original decision to access at $\bb$, hence avoiding collisions.


\subsection{Measurability}
Multiple protocols can be constructed from a stationary channel environment, due to the characteristic of RL, which causes non-static channel fluctuations and random realization of SDU arrivals. Thus, a method is needed to attain a consensus on which protocol should be used for the current environment. Inspired by the information theoretical metrics that evaluate classical protocols, we evaluate multiple SPMs based on semantic information theory. This is possible because SPM gains structural variability from the merging process of its vocabularies. On this note, we measure the level of randomness within a clause with the following definition of semantic entropy~\cite{Jinho_2022_SemCom}:
\begin{align} \label{eq:clause_Entropy}
    \mathcal{H}(c) = -\{c^\mathsf{P}\log(c^\mathsf{P}) + (1-c^\mathsf{P})\log(1-c^\mathsf{P}) \},
\end{align}where $c^\mathsf{P}=1$ for an extracted NPM, and $\mathcal{H}(c)$ is defined as $0$ for $c^\mathsf{P}=0$. Due to NPM's deterministic structure, it has a zero net semantic entropy. However, its structural complexity and excessive vocabularies due to the overparametrization makes it impractical to use. Consequently, extensive simulation runs are required to obtain an understanding of its operation or the task's complexity~\cite{Intrinsic_Dimention_ICLR2018}. In contrary, an SPM has a compact structure; and thanks to theoretical metrics, it can be evaluated without running costly and time-consuming simulation runs. For example, we can evaluate the net randomness within the SPM with the net model entropy:
\begin{align}
    \H_{net}(\hSPM)=\sum_{c \in \mathcal{S}} \mathcal{H}(c).
\end{align}
$\Hn$ is insightful due to its strong correlation to the UCM and DCM vocabulary sizes $|\vD|$ and $|\vU|$, respectively, and the downlink and action clause sets' cardinality $|B_i|$ and $|\Gamma_i|$, respectively. We can also calculate the partial net entropies for downlink clauses $\hb$ and action clauses $\hc$ by the following: $\H_{net}^\b (\hSPM)=\H_{\text{SPM}}(\hSPM \cap B)$, and $\H_{net}^\c (\hSPM)=\H_{\text{SPM}}(\hSPM \cap \Gamma)$. These metrics measure the net entropy pertaining to a specific clause type, which let us differentiate which logical connection type has more polysemous behavior within the protocol. For a consensual purpose, $\H_{net}^\b$ can be used by the BS to achieve a simpler control plane function, and $\H_{net}^\c$ can be used by the UEs to lower the variance of its actions. 
Lastly, by utilizing the SPM entropy, we can measure the task complexity by finding the minimum $\H_{net}(\hSPM^\psi)$ value among the protocols $\{\hSPM^1, ..., \hSPM^\psi\} $:
\begin{align}\label{eq:Complexity_condition}
    \min_\psi & \; \H_{net}(\hSPM^\psi).
\end{align}
The rationale behind this metric is that complex tasks require more vocabularies and connections to express the divergent situations present in the task. 

\subsection{Compactness}

So far, the communication environment parameters remained stationary, which we now relax and consider a non-stationary environment to address more realistic situations. Classical protocols are designed for a general purpose, so it may suffer from a non-stationary environment that deviates from the general purpose. On the other hand, NPM can be trained for a specific environment, but keeping and loading multiple models cause heavy memory overhead due to the overparametrization of NNs. In contrary, SPM's compactness makes it easy to store and compute. For example, consider an environment where UEs have periods of bursty SDUs, UEs leave and enter, or the wireless channel deteriorates over time. To cope with these non-stationary environment, keeping multiple compact SPMs as a portfolio grant model diversity and ensemble gain~\cite{Sejin_2020_uncertainty, ML_Diversity}.


\section{Simulation Results and Discussion} \label{Sec:Disc_Experim}

In this section, communication related hyperparameters are tuned to reflect a realistic communication scenario. We set $\lambda_1 = \lambda_2 =0.5$, $b_{\max}=5$, $D_{\max}=12$, $\rho_1=\rho_2 = 5$, $T=24$, and $\epsilon=0.02$ to model a situation where two UEs are contending for a noisy channel with equal SDU payload, and a BS capable of serving both UE's data stream successfully if the protocol is effective. We use the Pytorch library to implement NPMs, and we extend the Python ProbLog library to implement SPMs. The Huber loss function~\cite{huber_loss_1964} is chosen thanks to its robustness against outliers, reducing the fluctuation magnitude of DQN training; Adam optimizer is used with the initial learning rate of $0.0001$, the first and second exponential decay rate of $0.9$ and $0.999$, and epsilon as $10^{-7}$ for numerical stability. As additional KPIs to measure the contention performance, we denote $n_C$ and $n_D$ as the number of collisions and discarded SDUs, respectively. 
 
\subsection{Comparison with NPM and Slotted ALOHA} \label{subsec:Experiment}

\begin{figure*}[t]
  \centering  
  \includegraphics[height=11cm]{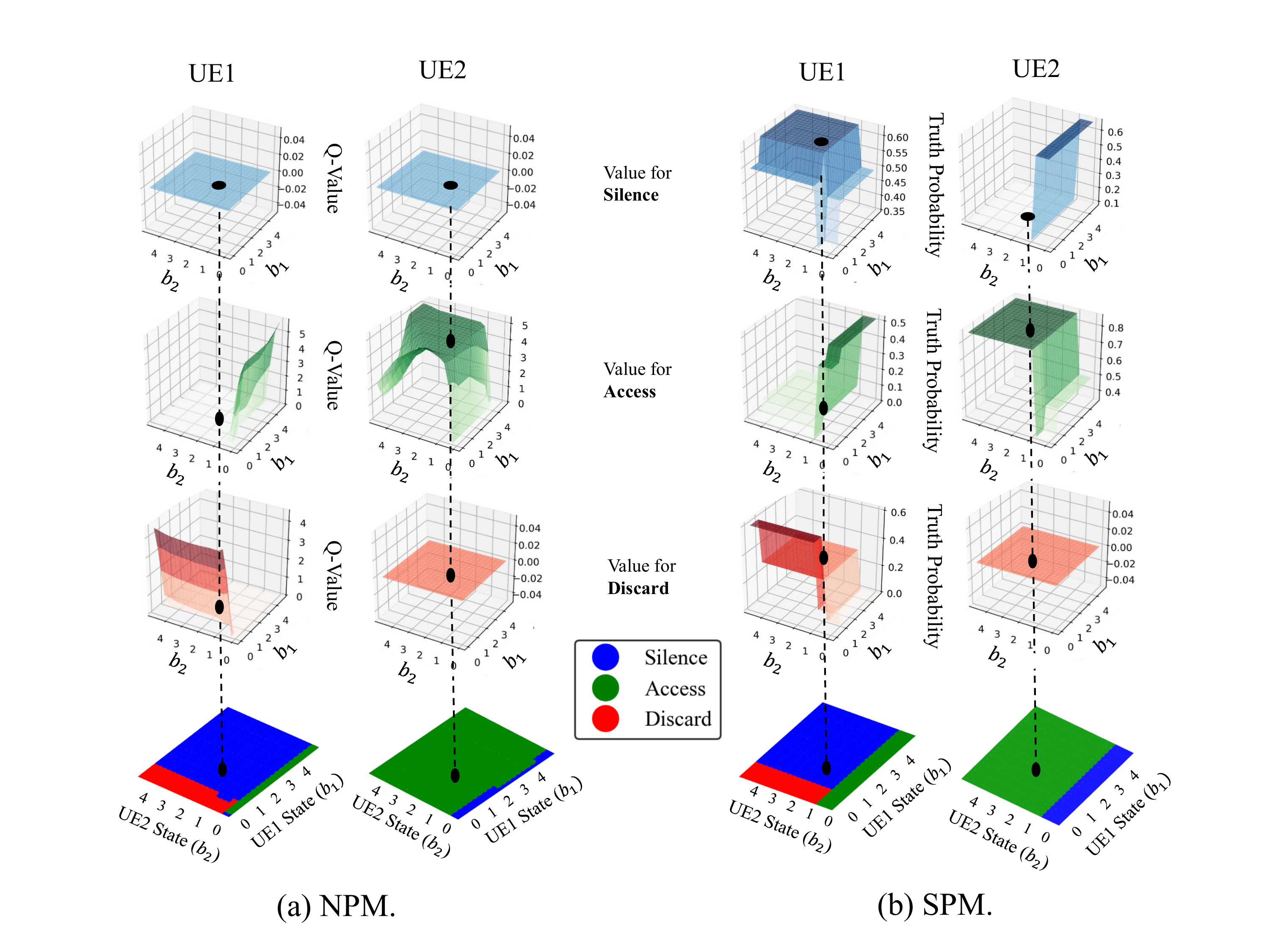}
  \caption{NPM vs. SPM: the operation results of UEs and the policy maps of (a) NPM, (b) SPM. Given a buffer state, NPM returns the Q-values for each action; then, the action with the highest Q-value is selected. An \stwo compares the truth probability of the actions given the input state, then chooses the action with the highest value. Plotting the decision regions for all input states gives the policy maps.} \label{Fig_NPMSPM}
\end{figure*} 

\begin{figure*}[t]
  \centering  
  \includegraphics[height=8cm]{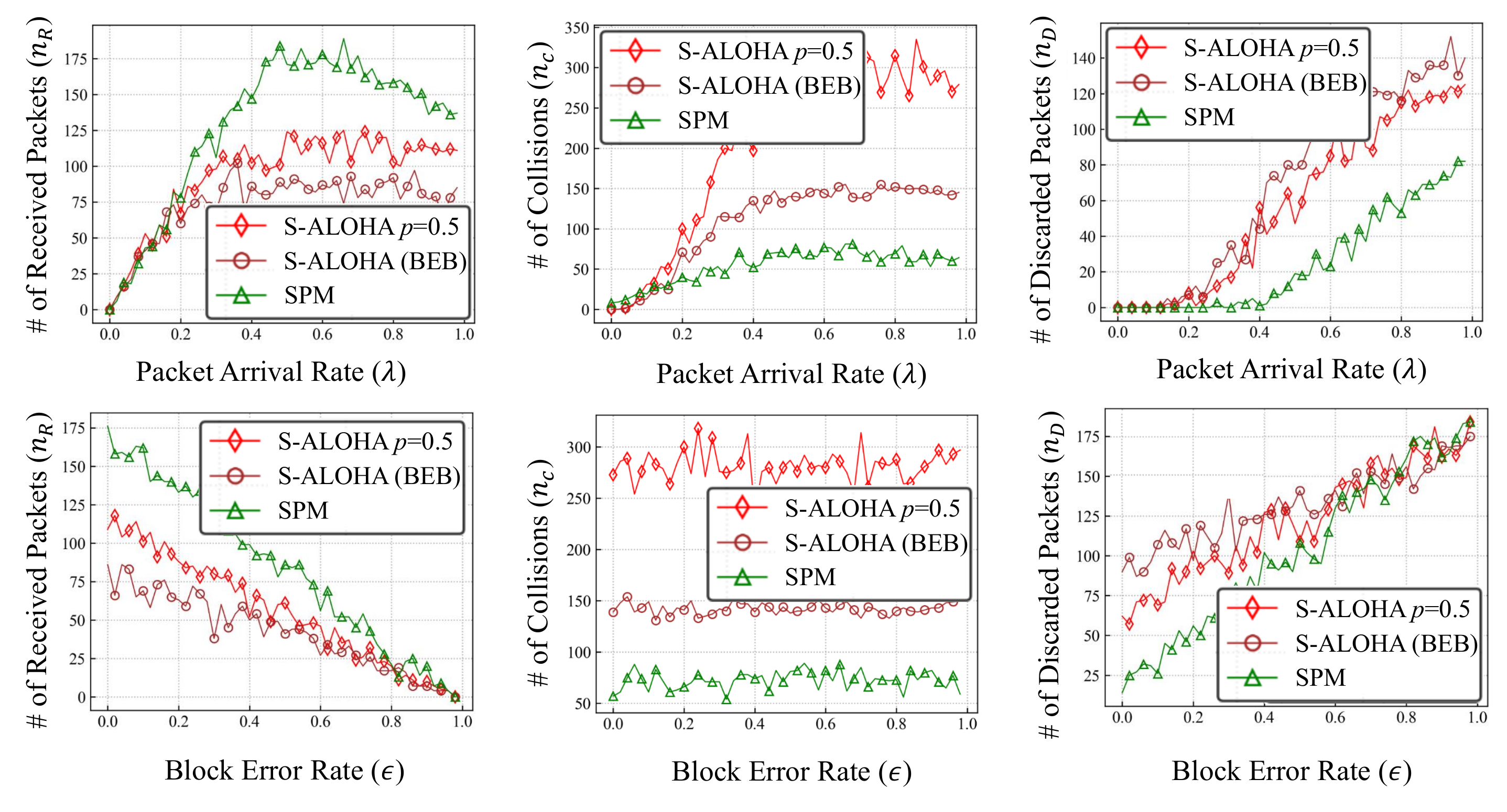}
  \caption{SPM vs. S-ALOHA: comparison of the contention performance of S-ALOHA, S-ALOHA with binary exponential backoff (BEB), and SPM: $n_C$, $n_R$, and $n_D$ at varied $\lambda$ and $\epsilon$.} \label{Fig_Aloha}
\end{figure*} 


Fig. \ref{Fig_NPMSPM} compares the emulation performance of an SPM. Fig. \ref{Fig_NPMSPM}(a) and (b) plot the second Q-value element that corresponds to the level of inclination of the UE to access the channel. Similarly, for an SPM, the truth probability of action $a_i$ displays the level of inclination to access, because it is the value that is compared with the other actions to decide whether to access or not. The Q-values are well imitated by the SPMs' truth probabilities for the access action in Fig. \ref{Fig_NPMSPM}(a) and (b). For $b_{\max}=5$, the policy emulation is correct for $97.22\%$ of the states. Minor deviation occurred when $b_1=b_2=0$, which is acceptable because the decision when both buffers are empty is negligible, which is corroborated by the identical average goodput performance $0.729$ at $\lambda=0.5$ and $\epsilon=0.02$.


In Fig. \ref{Fig_Aloha}, we check the contention performance of the SPMs compared to the conventional protocols. SPM's performance is compared with the performance of random access schemes such as slotted ALOHA (S-ALOHA) with access probability $p=0.5$~\cite{S_ALOHA_2002}, and S-ALOHA with binary exponential backoff (BEB) with base $2$ and adverse collision event. At $\lambda = 0.5$, SPM's $n_R$ is greater than S-ALOHA and S-ALOHA (BEB) by $182.2\%$ and $206.7\%$; and at $\epsilon=0.01$ by $161.5\%$ and $204.7\%$. This is due to the NPM counterpart's performance grounded on the environment, the SPM's intrinsic ability to emulate such behavior, and logical operations that obviate the polysemy problem. To verify the performance across diverse environments, we vary the packet arrival rate $\lambda$ and block error rate $\epsilon$. At each environment, the test is repeated $10$ times. The performance gap between SPM and random access schemes is evident except for when $\epsilon$ approaches $1$, where all SDUs are inevitably lost due to bad channel conditions. SPM shows improved performance regarding the KPIs $n_C$, $n_R$, and $n_D$ throughout other environments. Averaging over all environments, SPM's $n_R$ is greater than that of S-ALOHA and S-ALOHA (BEB) by $142.8\%$ and $184.4\%$; $n_D$ is reduced to $52.7\%$, $47.0\%$; and $n_C$ is reduced to $26.4\%$ and $50.9\%$.

\subsection{Collision-Free SPM} \label{subsec:Manipulation}

\begin{figure}[t]
  \centering
  \includegraphics[height=5cm]{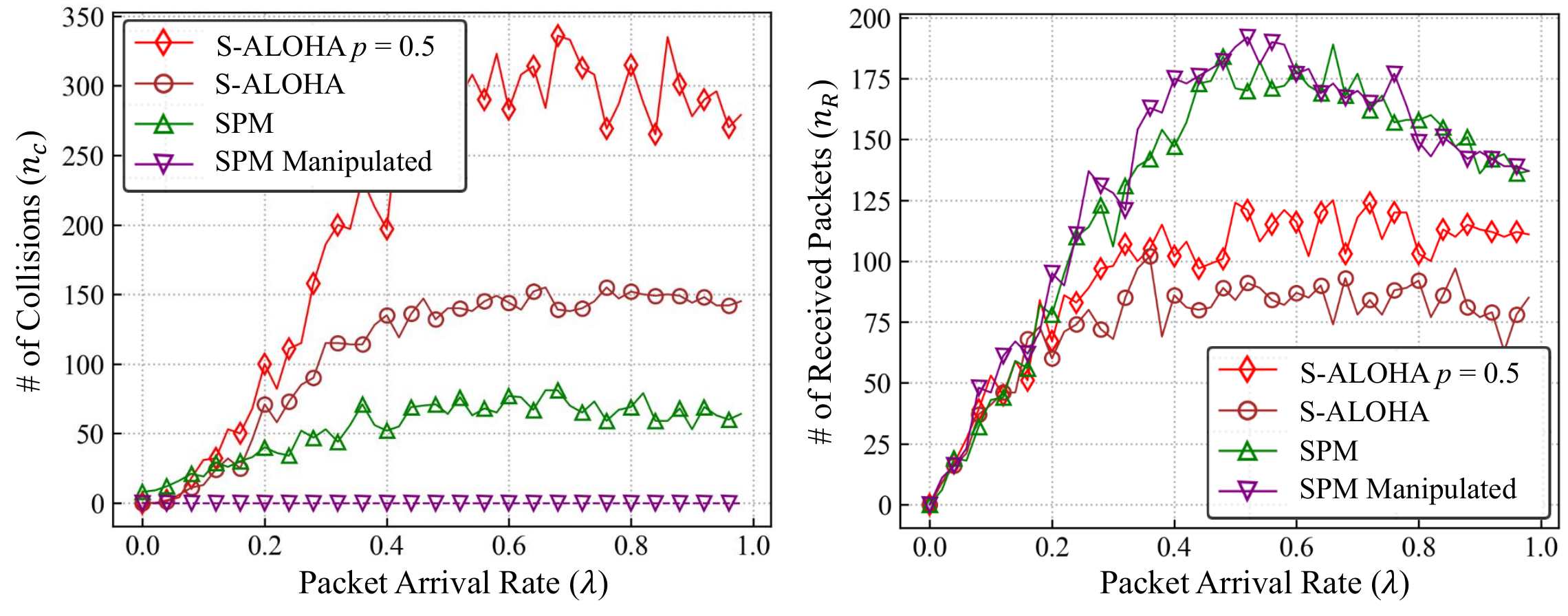}
  \caption{The results of manipulation for collision avoidance compared with the original SPM's and classical protocols' performance.}\label{Fig_Manip_CA}
\end{figure}

With simple manipulation steps, an SPM can be adjusted to perform well under a non-stationary environment where a new constraint such as collision avoidance should be considered.
Let $n_C^*$ be the number of collisions that the system should not exceed. For $\bb$, collision occurs when the actions $a_1^A$ and $a_2^A$ are chosen simultaneously, which occurs with the probability $\mathsf{Pr}(a_1^A, a_2^A|\hSPM,\bb) = \mathsf{Pr}(a_1^A |\hSPM,\bb)\mathsf{Pr}(a_2^A |\hSPM,\bb)$. We decide to manipulate $\hc$ for $\bb$ when $\mathsf{Pr}(a_1^A, a_2^A|\hSPM,\bb) > p_{th}$, where $p_{th}$ is the collision avoidance threshold set to satisfy $n_C^*$. To avoid collision for the input while making the least change to the protocol, we manipulate the SPM of the UE with lower access probability to remain silent instead, and choose an arbitrary UE for a tie. Suppose that UE $i$ has the lower access probability, we manipulate the $\hc_i^\mathsf{H}$ from $a_i^A$ to $a_i^S$ as explained in Section IV. Repeating this for all input states that exceeds $p_{th}$ ultimately manipulates an SPM to become collision-free. Fig. \ref{Fig_Manip_CA} shows the performance after manipulation for collision avoidance. With manipulations on actions that caused collisions, $n_C$ approaches $0$ for SPM after the manipulation. Only two steps of manipulation were required, which is much more efficient and scalable than manipulating an SPM, or retraining an NPM. The manipulation also resulted in a marginal improvement for $n_R$, i.e. $107.21\%$ improvement on average for all environments, which shows that the manipulation's net impact on the SPM was positive.

\subsection{Consensual SPM in a Stationary Environment}


\begin{figure}[t]
  \centering
  \subfigure[Protocol selection based on KPI.]{\includegraphics[height=5cm]{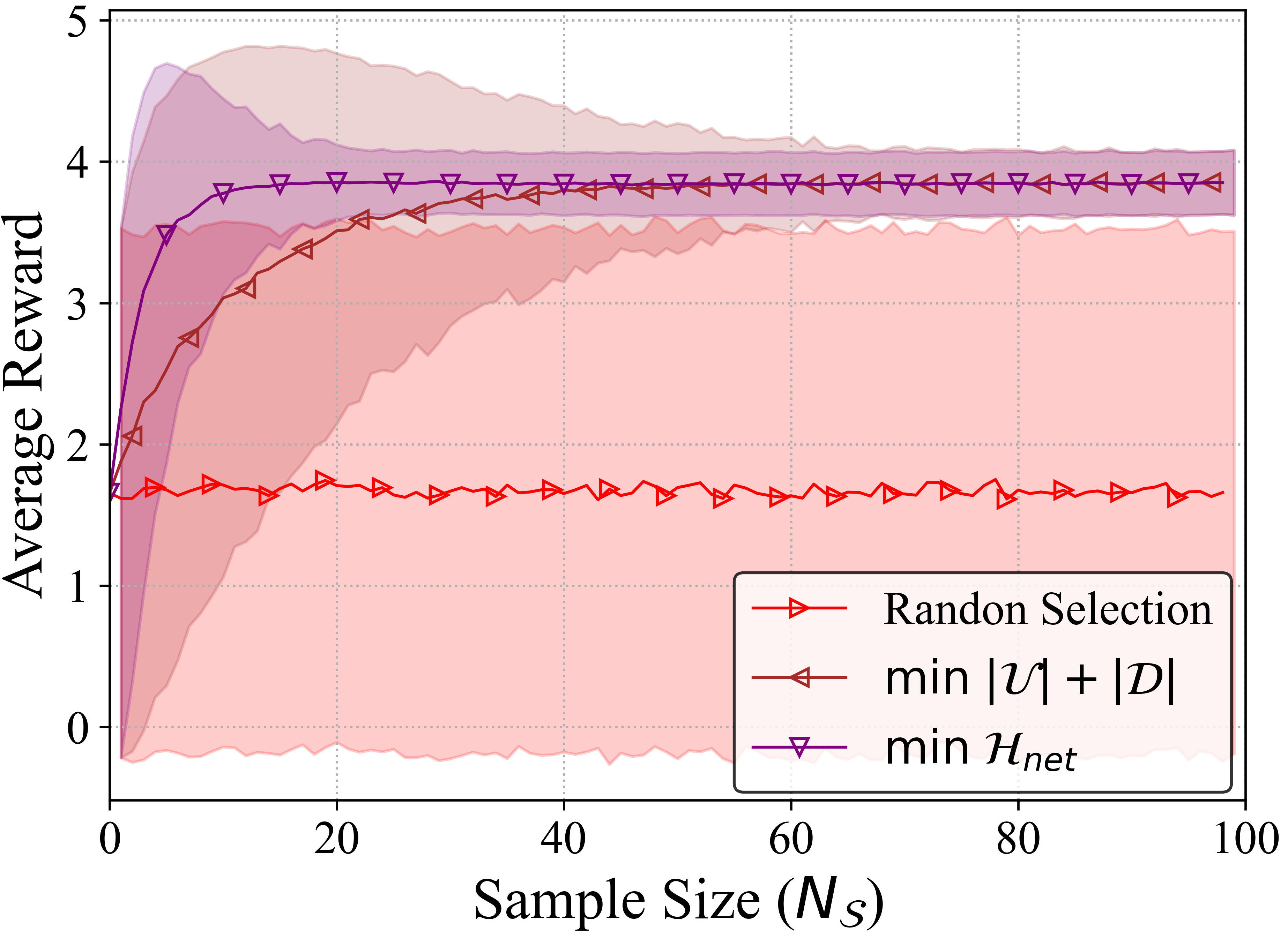}}
  \subfigure[Adaptation to non-stationary environment.]{\includegraphics[height=5cm]{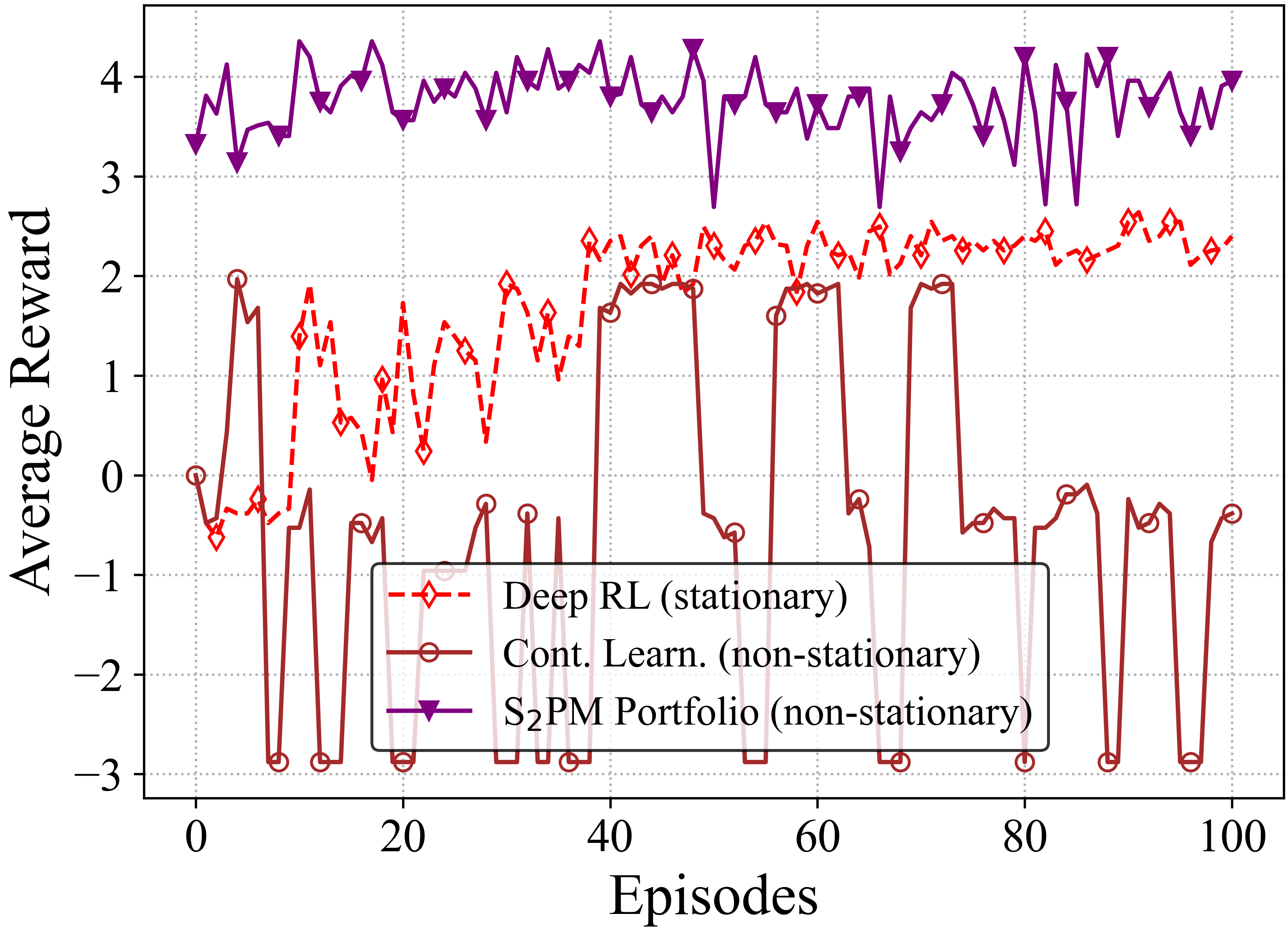}}
  \caption{(a) comparison of the average reward and its standard deviation for three selection schemes: choosing one \stwo randomly, choosing the \stwo with minimum net entropy, and choosing the \stwo with the least number of vocabularies. (b) comparison of the average reward of deep RL in a stationary environment, and two schemes in a non-stationary environment, which are continual learning and \stwo portfolio.} \label{Fig_SPM_Sel}
\end{figure}

Based on the theoretical evaluation tools from Sec. \ref{Sec:Potentials}, we can select one model among multiple models with simple evaluation on the protocol structure, without running performance tests. Fig. \ref{Fig_SPM_Sel}(a) compares the average rewards of three different selection schemes. Given a set of $N_\SPM$ \stwo samples, the random selection scheme chooses one \stwo randomly from the set; $\min \; |\vU| +|\vD|$ scheme chooses the \stwo that uses the least UCM and DCM vocabularies; and the $\min\ \H_{net}$ scheme chooses the \stwo with the smallest net entropy. In the experiment, we experiment with $200$ SPMs, and $N_\SPM$ \stwon s are sampled from them. $3000$ trials are repeated at each $N_\SPM$. The results show that the net SPM entropy is the best metric for getting a consensual \stwo among the three, because it selects the well-performing \stwo from smaller sample sizes. Selecting according to the minimum net entropy gives an average reward that converges to $3.84$ around $N_\SPM=20$ with the standard deviation of $0.22$, and choosing for the minimum vocabulary size gives an average reward that converges to $3.84$ around $N_\SPM=60$, whereas random selection scheme's average reward stays between $1.5$ and $1.8$ for all sample sizes, with the average standard deviation of $1.86$. 


\subsection{Portfolio SPM for a Non-Stationary Environment}\label{subsec:Memory}

Thanks to the compactness of SPMs, we can maintain a portfolio of SPMs for a non-stationary environment where communication parameters change frequently. 
To verify the robustness of an SPM portfolio, we compare it with a continual learning based NPM over a changing environment. Fig. \ref{Fig_SPM_Sel}(b) compares two schemes that try to adapt to a frequently changing environment, and an NPM that is trained for a stationary environment, to provide a baseline. In the simulation, we try to model a non-stationary environment with UEs that have incoming bursts of SDUs by a two-state Markov chain, where in one state UE $1$ has $\lambda = 0.9$ and UE $2$ has $\lambda = 0.1$, and in the other state, the packet arrival rates are reversed; the state transition probability is $0.8$ for both states. In such a dynamic environment, a protocol that prioritizes only one UE to transmit regardless of the other UE suffers when the environment state is changed. The red triangle graph shows how the training would have progressed if the environment was stationary, with each UE having $\lambda=0.5$; the blue circled graph shows the test performance of a protocol being adjusted by continual learning, which retrains at the current environment, starting from the NPM of the previous state. The purple diamond graph shows the test performance of an SPM portfolio, with SPMs constructed for the two alternating environments. From the figure, it can be observed that continual learning leads to detrimental results, i.e. average reward being less than $0$ for $74.3\%$ of the communication cycles, which is due to catastrophic forgetting~\cite{Catastrophic_1999}, and learning being stuck at bad local minima when the environment changes frequently. In contrast, an SPM portfolio never falls below the average reward of $2.63$ because the portfolio contains a model that is constructed for each environment.

\section{Conclusion}

In this work, we proposed a novel MAC protocol model extracted and symbolized from an NN-based NPM, coined an SPM. The SPM turns the black-box NN into a directed graph or a collection of semantic clauses written in ProbLog, a logic programming language for symbolic AI. The SPM is therefore interpretable by both humans and machines, and instantly reconfigurable. Furthermore, the SPM is measurable via semantic entropy, and compact occupying only $0.02$\% of the memory compared to its original SPM, thereby enabling best SPM selection and SPM portfolio applications.
There are several intriguing future problems such as extensions to dynamic spectrum access, viewing this problem from a privacy standpoint, and exploiting SPMs for CM error correction. Furthermore, while we focus only on adjusting SPMs, it could be an interesting topic to jointly design NPM learning and SPM construction.







\bibliographystyle{ieeetr}
\bibliography{semantic}

\vspace{12pt}

\end{document}